\documentclass[aps, prb, twocolumn, superscriptaddress, longbibliography, floatfix]{revtex4-2}
\usepackage{amsmath}
\usepackage{amssymb}
\usepackage{graphicx}
\usepackage{amsmath}
\usepackage{amssymb}
\usepackage{amsfonts}
\usepackage{color,xcolor}

\usepackage{xfrac}
\usepackage{ulem}

\begin{document}
\title{Quasimolecular electronic structure of the spin-liquid candidate Ba$_3$InIr$_2$O$_9$}

\author{A. Revelli}
\affiliation{Institute of Physics II, University of Cologne, 50937 Cologne, Germany}
\author{M. Moretti Sala}
\affiliation{European Synchrotron Radiation Facility, BP 220, F-38043 Grenoble Cedex, France}
\affiliation{Dipartimento di Fisica, Politecnico di Milano, I-20133 Milano, Italy}
\author{G. Monaco}
\affiliation{Dipartimento di Fisica e Astronomia "Galileo Galilei", Universit\`{a} di Padova, Padova, Italy}
\author{M. Magnaterra}
\affiliation{Institute of Physics II, University of Cologne, 50937 Cologne, Germany}
\author{J. Attig}
\author{L. Peterlini}
\affiliation{Institute for Theoretical Physics, University of Cologne, 50937 Cologne, Germany}
\author{T.~Dey}
\affiliation{Institute of Physics II, University of Cologne, 50937 Cologne, Germany}
\affiliation{Department of Physics, IIT (ISM) Dhanbad,  Jharkhand 826004, India}
\affiliation{Experimental Physics VI, Center for Electronic Correlations and Magnetism, University of Augsburg, 86159 Augsburg, Germany}
\author{A.A.~Tsirlin}
\affiliation{Experimental Physics VI, Center for Electronic Correlations and Magnetism, University of Augsburg, 86159 Augsburg, Germany}
\author{P.~Gegenwart}
\affiliation{Experimental Physics VI, Center for Electronic Correlations and Magnetism, University of Augsburg, 86159 Augsburg, Germany}
\author{T.~Fr\"{o}hlich}
\author{M.~Braden}
\author{C.~Grams}
\author{J.~Hemberger}
\affiliation{Institute of Physics II, University of Cologne, 50937 Cologne, Germany}
\author{P. Becker}
\affiliation{Sect.\ Crystallography, Institute of Geology and Mineralogy, University of Cologne, 50674 Cologne, Germany}
\author{P.H.M.~van Loosdrecht}
\author{D.I. Khomskii}
\affiliation{Institute of Physics II, University of Cologne, 50937 Cologne, Germany}
\author{J. van den Brink}
\affiliation{Institute for Theoretical Solid State Physics, IFW Dresden, 01069 Dresden, Germany}
\affiliation{Institute for Theoretical Physics and W\"urzburg-Dresden Cluster of Excellence ct.qmat, Technische Universit\"at Dresden, 01069 Dresden, Germany}
\author{M.~Hermanns}
\affiliation{Department of Physics, Stockholm University, AlbaNova University Center, SE-106 91 Stockholm, Sweden}
\affiliation{Nordita, KTH Royal Institute of Technology and Stockholm University, SE-106 91 Stockholm, Sweden}
\author{M. Gr\"{u}ninger}
\affiliation{Institute of Physics II, University of Cologne, 50937 Cologne, Germany}

\date{July 11, 2022; revised: September 21, 2022}

\begin{abstract}
The mixed-valent iridate Ba$_3$InIr$_2$O$_9$ has been discussed as a promising candidate for
quantum spin-liquid behavior. The compound exhibits Ir$^{4.5+}$ ions in face-sharing 
IrO$_6$ octahedra forming Ir$_2$O$_9$ dimers with three $t_{2g}$ holes per dimer.
Our results establish Ba$_3$InIr$_2$O$_9$ as a cluster Mott insulator.
Strong intradimer hopping delocalizes the three $t_{2g}$ holes in quasimolecular dimer states
while interdimer charge fluctuations are suppressed by Coulomb repulsion.
The magnetism of Ba$_3$InIr$_2$O$_9$ emerges from spin-orbit entangled quasimolecular moments
with yet unexplored interactions, opening up a new route to unconventional magnetic properties 
of $5d$ compounds. Using single-crystal x-ray diffraction we find the 
monoclinic space group $C2/c$ already at room temperature. 
Dielectric spectroscopy shows insulating behavior.
Resonant inelastic x-ray scattering (RIXS) reveals a rich excitation spectrum below $1.5$\,eV 
with a sinusoidal dynamical structure factor that unambiguously demonstrates the 
quasimolecular character of the electronic states. Below 0.3\,eV, we observe a 
series of excitations. According to exact diagonalization calculations, such low-energy 
excitations reflect the proximity of Ba$_3$InIr$_2$O$_9$ to a hopping-induced phase transition 
based on the condensation of a quasimolecular spin-orbit exciton. 
The dimer ground state roughly hosts two holes in a bonding $j$\,=\,1/2 orbital and 
the third hole in a bonding $j$\,=3/2 orbital.
\end{abstract}

\maketitle

\section{Introduction}

The spin-orbit-assisted Mott-insulating character of iridates provides a very active 
playground for quantum magnetism \cite{WitczakKrempa14,Rau16,Schaffer16,Trebst17,Winter17,Hermanns18,Cao18,Takagi19,Motome20,Takayama21,Nguyen21,Trebst22}.
For $5d^5$ and $5d^4$ configurations, spin-orbit coupling $\lambda$\,=\,$\zeta/2S$ lifts the 
degeneracy of the $t_{2g}$ orbitals and yields electronic bands which in many compounds are 
sufficiently narrow for a moderate on-site Coulomb interaction $U$ to result in Mott insulators 
with local $j$\,=\,$1/2$ and $j$\,=\,$0$ moments, respectively.
Particular attention has been paid to $j$\,=\,$1/2$ moments in compounds with corner-sharing or edge-sharing
IrO$_6$ octahedra, such as Sr$_2$IrO$_4$ or Na$_2$IrO$_3$ with dominant isotropic Heisenberg exchange or
bond-directional Kitaev exchange couplings, respectively \cite{Jackeli09}.
The experimental realization of Kitaev couplings has triggered an intense quest for novel quantum spin-liquid
states in so-called Kitaev materials \cite{Trebst17,Singh10,Singh12,Kitagawa18,Takagi19,Motome20,Trebst22}.

\begin{figure}[tb]
\centering
\includegraphics[width=0.4\columnwidth]{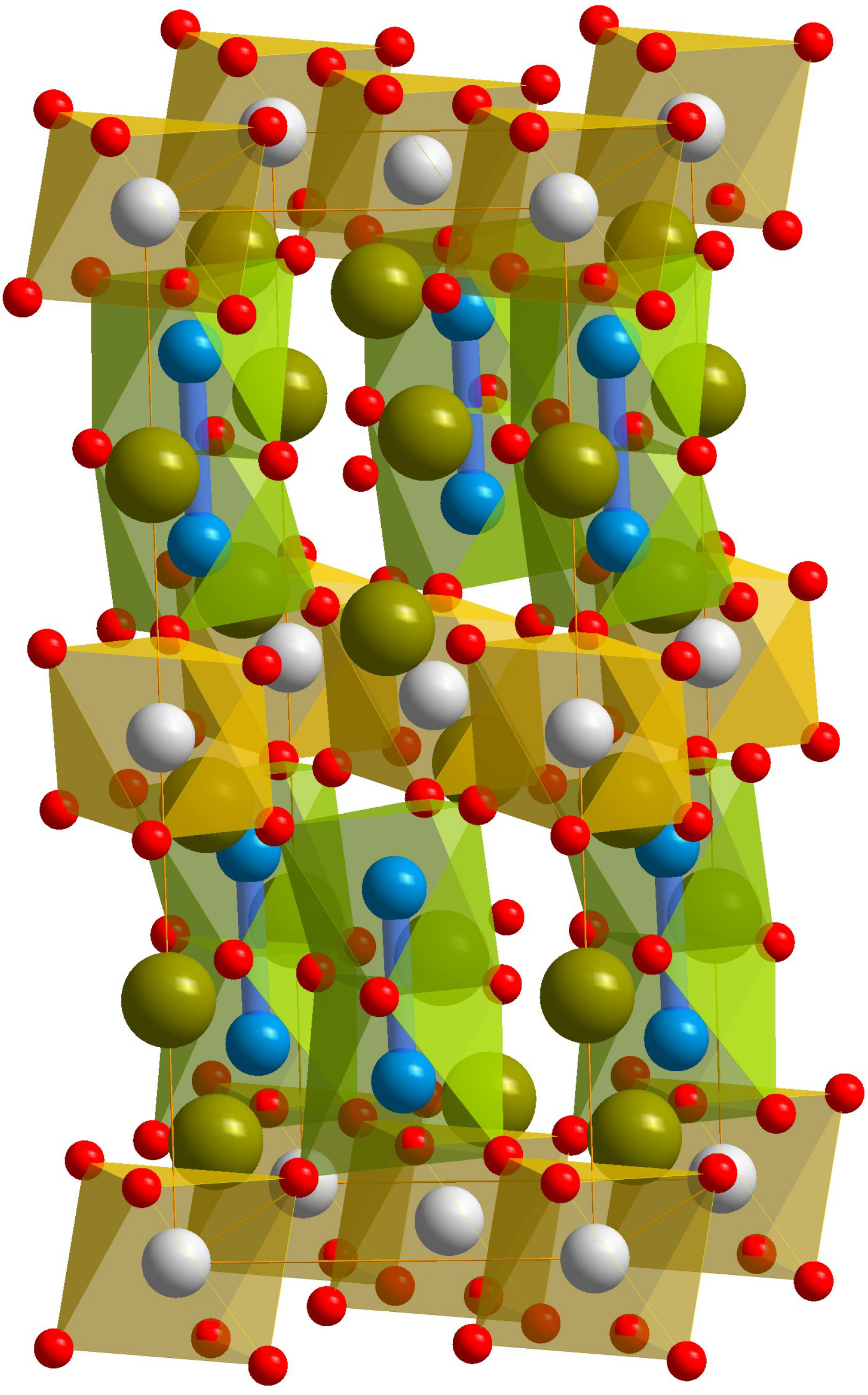}
\includegraphics[width=0.08\columnwidth]{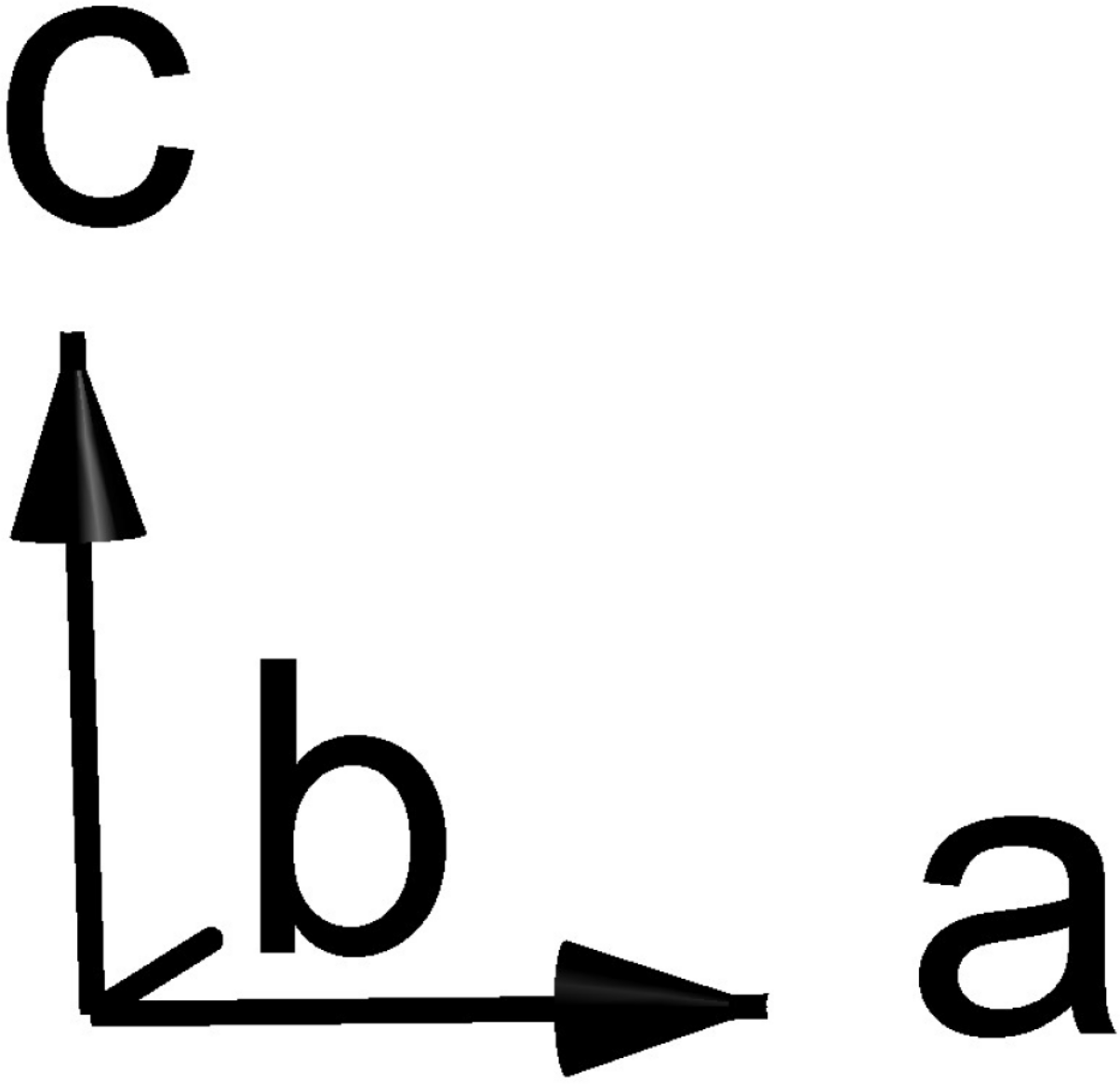}
\includegraphics[width=0.44\columnwidth]{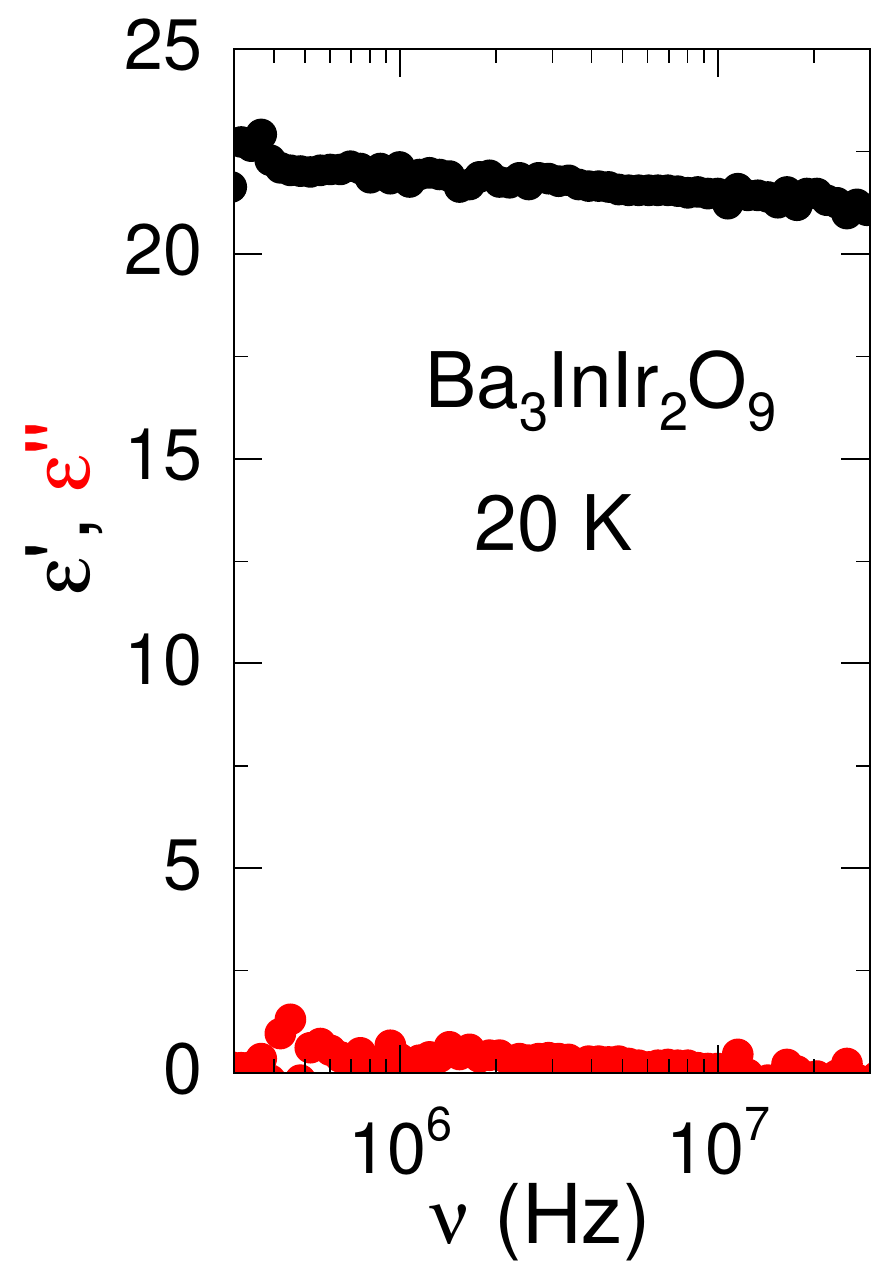}
   \caption{Left: Sketch of the crystal structure of Ba$_3$InIr$_2$O$_9$ with triangular layers of 
   	well-separated Ir$_2$O$_9$ units. Bold blue lines connect Ir ions within a dimer.
   	Olive/gray/red: Ba/In/O. 
   	Right: Permittivity $\varepsilon^{\prime}$ (black) and dielectric loss $\varepsilon^{\prime\prime}$ (red)
	at 20\,K in the microwave range. The small $\varepsilon^{\prime\prime}$ demonstrates the
	insulating character of Ba$_3$InIr$_2$O$_9$.
}
\label{fig:struc}
\end{figure}

The rich physics encountered in the iridates is based on the subtle balance of $U$, $\zeta$, and hopping $t$.
Compared to corner- or edge-sharing IrO$_6$ octahedra, the less intensively studied case of
\textit{face-sharing} octahedra features enhanced hopping \cite{KhomskiiZhETF},
opening access to unexplored regions in $U$-$\zeta$-$t$ space.
We focus on the structural motif of two face-sharing octahedra that form Ir$_2$O$_9$ units or dimers,
as realized in the Ba$_3$$X$Ir$_2$O$_9$ family with the layered
($6H$)-BaTiO$_3$ structure formed for a wide variety of mono-, di-, tri-, or tetravalent $X$
ions \cite{Doi04,Kim04,Sakamoto06,Nguyen21,Dey12,Dey14,Kumar16,Lee17,Dey17,Khan19,Nag19,Kumar21,Garg21}, 
see Fig.\ \ref{fig:struc}.
Within the layers, the Ir$_2$O$_9$ units are arranged on a triangular lattice.
In total there are 12 neighbors in three dimensions, analogous to a \textit{hcp} lattice of Ir$_2$O$_9$ units.
In Ba$_3$InIr$_2$O$_9$ with Ir$^{4.5+}$ ions and three holes per dimer, no long-range magnetic order
is observed but persistent spin dynamics were reported down to 20\,mK using thermodynamic measurements,
nuclear magnetic resonance, and muon spin resonance \cite{Dey17}. This promisingly points toward a
possible gapless spin-liquid ground state.

\begin{figure}[tb]
	\centering
	\includegraphics[width=0.84\columnwidth]{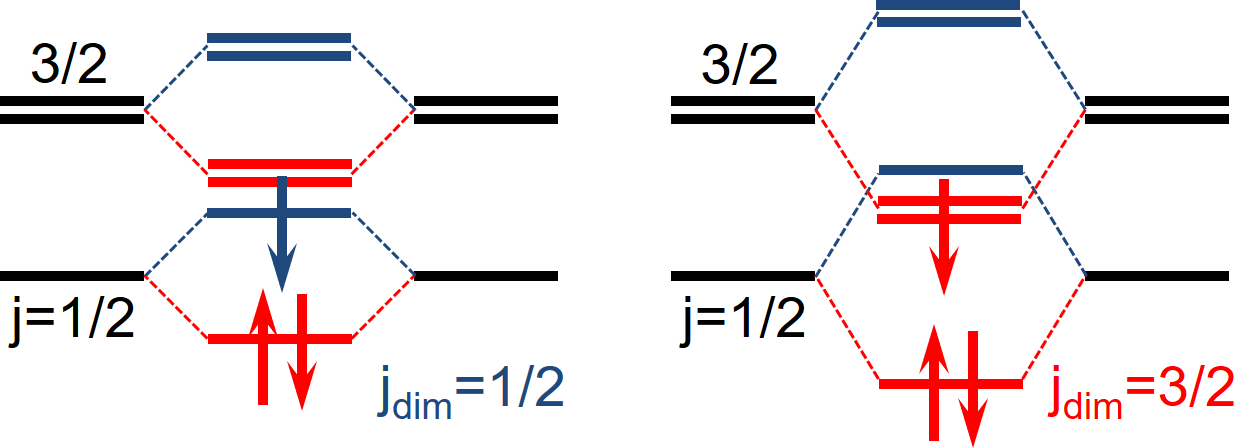}
	\caption{Quasi-molecular orbitals based on local single-hole states with $j$\,=\,1/2 
	and 3/2. In the hole picture, bonding (antibonding) orbitals are depicted in red (blue). 
	In this $j$ basis, hopping is block-diagonal for $t_{e_g^\pi}$\,=\,$-t_{a_{1g}}$ 
	(see main text). For three holes, increasing hopping yields a phase transition 
	from $j_{\rm dim}$\,=\,1/2 (left) to $j_{\rm dim}$\,=\,3/2 (right) \cite{Li20}. 
	In the latter state, all three holes occupy bonding orbitals. 
	In Ba$_3$InIr$_2$O$_9$, the hopping parameters deviate from the idealized case $t_{e_g^\pi}$\,=\,$-t_{a_{1g}}$ depicted here, but this sketch still provides 
	a good starting point for an intuitive understanding of the electronic structure. 
	}
	\label{fig:sketch}
\end{figure}

An essential step for the understanding of the magnetism of Ba$_3$$X$Ir$_2$O$_9$  
is the determination of the character of the magnetic moments.
To this end, one has to resolve the electronic structure of the Ir$_2$O$_9$ units.
In the limit of dominant $U$, one expects spin-orbit entangled moments localized 
on each Ir site with strong exchange interactions \cite{Xu19}.
For mixed-valent compounds such as Ba$_3$InIr$_2$O$_9$, this limit may, e.g., feature charge ordering of 
$5d^5$ Ir$^{4+}$ and $5d^4$ Ir$^{5+}$ sites with $j$\,=\,$1/2$ and $j$\,=\,$0$ moments coexisting 
on a dimer. However, depending on the choice of $X$ ions the Ir$_2$O$_9$ units exhibit a 
very short Ir-Ir distance of 2.5--2.6\,\AA\ \cite{Doi04,Dey17}, shorter than the value realized 
in Ir metal, 2.7\,\AA. This suggests an alternative scenario in which the large intradimer 
Ir-Ir hopping yields quasimolecular orbitals with possibly quenched orbital moments \cite{Streltsov16}, 
as proposed for Na$_2$IrO$_3$ \cite{Mazin12,Foy13}. 
For iridate dimers, the formation of quasimolecular orbitals was claimed in 
Ba$_5$AlIr$_2$O$_{11}$, Ba$_3$BiIr$_2$O$_9$, and Ba$_3$CeIr$_2$O$_9$ 
\cite{Streltsov17,Wang18,Miiller12,Revelli19}.
For the latter compound, it has been shown that RIXS measurements of the dynamical 
structure factor are particularly well suited to nail down a quasimolecular orbital 
character of the electronic states as well as to unravel the role of 
spin-orbit coupling \cite{Revelli19}.
In Ba$_3$CeIr$_2$O$_9$ with two holes per dimer, the comparison of RIXS and theory established 
that the quasimolecular states are predominantly built from spin-orbit entangled $j$\,=\,$1/2$ 
states, giving rise to a pseudospin $j_{\rm dim}$\,=\,$0$ character as opposed to a spin-singlet 
dimer \cite{Revelli19}.
The compound Ba$_3$InIr$_2$O$_9$ with three holes per dimer has been discussed as being 
close to a phase transition \cite{Li20} between a $j_{\rm dim}$\,=\,1/2 state 
with dominant spin-orbit coupling and a $j_{\rm dim}$\,=\,3/2 state for larger hopping, 
see Fig.\ \ref{fig:sketch}.
This phase transition is related to van-Vleck-type excitonic magnetism \cite{Khaliullin13}, 
also known as singlet magnetism \cite{KhomskiiBook}, 
in nominally nonmagnetic $d^4$ $j$\,=\,0 compounds in which magnetic
moments arise from the hopping- or exchange-induced condensation of excited states. 
Similarly, the condensation of spin-orbit excitons has been discussed in one-dimensional 
$t_{2g}$ systems \cite{Kaushal19}.  
The quasimolecular dimer orbitals provide a different flavor of this mechanism. 
Condensation of the $j_{\rm dim}$\,=\,3/2 excitation occurs on the 
dimer level based on strong intradimer hopping, while exchange between the 
dimer moments is related to interdimer hopping with a smaller energy scale.

Here, we study the crystal structure and the electronic structure of Ba$_3$InIr$_2$O$_9$, 
which represents a model system for Ir dimers with unpaired electrons. We find a rich RIXS 
spectrum of intra-$t_{2g}$ excitations and demonstrate their quasimolecular nature by the 
observation of a sinusoidal RIXS interference pattern, i.e., the dynamical structure factor
of a dimer, similar to the two-hole case of Ba$_3$CeIr$_2$O$_9$ \cite{Revelli19}.
A comparison of RIXS spectra with exact diagonalization calculations establishes a 
$j_{\rm dim}$\,=\,$3/2$ nature of the ground state, 
while the observation of RIXS features below about 0.3\,eV supports that Ba$_3$InIr$_2$O$_9$ 
is close to the phase transition to a $j_{\rm dim}$\,=\,$1/2$ state.

The existence of quasimolecular magnetic moments in dimers with an odd number of charge 
carriers clearly establishes that Ba$_3$InIr$_2$O$_9$ belongs to the class of cluster Mott 
insulators \cite{AbdElmeguid04,Chen14,Khomskii21}.
Due to strong intradimer Ir-Ir hopping, charge carriers are delocalized inside the 
dimer-type cluster in quasimolecular orbitals. The Mott gap in this case refers to 
\textit{inter}-dimer charge excitations which are suppressed by Coulomb repulsion.
In comparison, the two holes per dimer in the sister compound Ba$_3$CeIr$_2$O$_9$ fully 
occupy the lowest-lying molecular orbital, leading to a nonmagnetic ground state. 
Cluster Mott insulators offer a new route to tailor spin-orbit-entangled quasimolecular 
magnetic moments. In this spirit, our RIXS results reveal the local quasimolecular orbital 
structure of the spin-liquid candidate material Ba$_3$InIr$_2$O$_9$, providing an 
essential step toward a thorough understanding of its magnetic properties.

\section{Experimental}
\label{sec:exp}

Single crystals of Ba$_3$InIr$_2$O$_9$ were grown at 1200$^\circ$C in air using a BaCl$_2$ flux, 
as described in Ref.\ \cite{Li20}.
We first synthesized polycrystalline material using stoichiometric amounts of In$_2$O$_3$, BaCO$_3$,
and Ir powder \cite{Dey17}. This polycrystalline material was used to grow the single crystals.
Single-crystal x-ray diffraction measurements were carried out using a Bruker AXS Kappa APEX II
four-circle diffractometer that operates with a wave length of $0.7\,$\AA\ (Mo $K_{\alpha}$ radiation).
The studied crystal was shaped like a hexagonal prism and a numerical absorption correction is 
based on a model with eight faces at distances from the center of about 100\,$\mu$m.

RIXS measurements were performed at beamline ID20 at the ESRF \cite{Moretti18}.
The crystal studied in RIXS had the shape of a hexagonal prism with a size of about 0.3\,mm
and a thickness of roughly 0.05\,mm.
The RIXS intensity of intra-$t_{2g}$ excitations was resonantly maximized by choosing an incident energy
of 11.215\,keV at the Ir $L_3$ edge.
Data were collected at 10\,K.\@
At low temperature, a mono\-clinic distortion with a monoclinic angle $\beta$\,=\,$90.854(3)^\circ$
was reported based on powder neutron diffraction data \cite{Dey17}.
Orienting the sample on the RIXS spectrometer with a spot size of the incident beam of about
$(20\times 10)\,\mu$m$^2$, we found that the measurement averages over different monoclinic domains.
With $\sin\beta \! \approx \! 0.9999$, we neglect the small distortion for the sake of simplicity
and assume hexagonal $P6_3/mmc$ symmetry for the RIXS measurements.
We studied the (001) surface in horizontal scattering geometry with the \textit{c} axis in the
scattering plane and the \textit{a} axis along the vertical direction.
The incident photons were $\pi$ polarized.
Combining a Si(844) backscattering monochromator and $R$\,=\,2\,m Si(844) spherical diced crystal
analyzers, an overall energy resolution of 25\,meV was achieved \cite{Moretti13}.
Replacing the Si(844) backscattering monochromator with a Si(311) channel-cut yields a lower resolution
of 0.36\,eV but a larger signal-to-noise ratio.
RIXS spectra were measured with the high-resolution setup by scanning the energy loss at constant
transferred momentum {\bf q}. 
The spectra were normalized by the incident flux and the acquisition time. 
Additionally, we collected {\bf q} scans at constant energy loss,
both with the low- and the high-resolution setup.
For comparison, we measured RIXS on Ba$_3$CeIr$_2$O$_9$ using the same setup and the same sample 
orientation as for Ba$_3$InIr$_2$O$_9$, as described previously \cite{Revelli19}.

\section{Results}

\subsection{Insulating behavior}

We probe the conductivity of Ba$_3$InIr$_2$O$_9$ by dielectric spectroscopy in the microwave range. 
A 4-point measurement of the dc resistivity was unfeasible since the size of our single crystals 
does not exceed 0.3\,mm. 
We employ a 2-point approach in which case extrinsic contact effects have to be considered.
These can be reduced by measuring at frequencies that are high enough
to short out the contacts.
The crystal was contacted in capacitor geometry on a microstrip waveguide with the electric
field parallel to the $c$ axis.
Using a vector network analyzer (Rhode\&Schwarz ZNB8) in the frequency range from 300\,kHz to 30\,MHz,
the permittivity of Ba$_3$InIr$_2$O$_9$ was measured at 20\,K, see Fig.\ \ref{fig:struc}.
A significant uncertainty in the absolute values is caused by the high relative error
in the determination of the small effective area $A_s$\,=\,0.02\,mm$^2$ and 
thickness $d_s$\,=\,0.04\,mm.
However, the experimental result corresponds to a real part of the conductivity
$\sigma \approx 10^{-6}/\Omega$\,cm,
which firmly establishes the insulating character of Ba$_3$InIr$_2$O$_9$.
Independent of the exact absolute values, this result is supported by the observation that
the dielectric loss $\varepsilon^{\prime\prime}$ is much smaller than the permittivity $\varepsilon^{\prime}$,
as expected for an insulator.

\subsection{Single-crystal structure refinement}
\label{sec:struc}

In single-crystal x-ray diffraction at 295\,K, we collected the intensities of $69\,971$ Bragg
reflections, yielding $877$ independent reflections with respect to the hexagonal space group $P 6_3/m m c$.
The integration yields an internal $R$ value \cite{karplus}
of $R^2$(int)\,=\,6.53\,\% after absorption correction.
Using the software JANA \cite{petricek}, three refinements were carried out in space group $P 6_3/m m c$.
One assumes the ordered structure, with In (Ir) atoms located on the Wyckoff sites 2\textit{a} (4\textit{f}).
The second refinement considers a statistical distribution of In and Ir atoms,
with the 2\textit{a} and 4\textit{f} sites both occupied by 1/3 In and 2/3 Ir.
The third refinement assumes partial disorder, allowing free occupations of these sites
with the constraint that each site is fully occupied.
We employ isotropic atomic displacement parameters for all atoms and Gaussian isotropic
Becker-Coppens extinction correction \cite{Becker1974a}. For the refinements with In and Ir
sharing a site, these atoms are constrained to have the same atomic displacement parameters.

The best result was obtained for the partially ordered model. It yields
$R$(obs)\,=\,5.96\,\%, w$R$(obs)\,=\,8.41\,\%, $R$(all)\,=\,5.96\,\%, 
and w$R$(all)\,=\,8.41\,\%.
We find that the 2\textit{a} site is occupied by 98.5(1.1)\,\% In atoms and 
1.5(1.1)\,\% Ir atoms, while the 4\textit{f} sites are occupied by 
92.7(1.4)\,\% Ir atoms and 7.3(1.4)\,\% In atoms.
For comparison, the fully disordered model can be ruled out, w$R$(all)\,=\,19.56\,\%,
while the ordered model yields w$R$(all)\,=\,9.14\,\%.
In total, the single-crystal data point toward a small surplus of In ions with 
about 7\% of In ions on the Ir 4\textit{f} dimer sites while the In 2\textit{a} sites 
are almost completely occupied by In. This has to be compared with the powder data 
which indicate In-Ir disorder of about 2.8(5)\% \cite{Dey17} in a refinement
with fixed In to Ir total occupation ratio.

We have searched for a breaking of the translational symmetry by analyzing half-integer 
indexed Bragg reflections with respect to the hexagonal lattice.
In standard operation of the x-ray generator with 
50\,kV voltage many of these reflections were observed with statistically significant intensities.
At 30\,kV, which is below the voltage required to generate photons with half wavelength, 0.35\,\AA , 
these signals are suppressed although we significantly enhanced the counting time to 2000\,s/degree. 
All half-indexed reflections can thus be attributed to $\lambda/2$ contamination that possesses a greater
impact in x-ray diffraction experiments on highly absorbing materials like iridates.
The comparison of all observed half-indexed reflections with their doubled integer-indexed partners
further corroborates this conclusion.

\begin{table}[t]
	\begin{tabular}[t]{cp{.2mm}cp{.2mm}cp{.2mm}cp{.2mm}cp{.2mm}c}  %p{2mm}
		& &     x            & &     y       & &     z         & &  U$_{\rm iso}$  & &     occ. \\\hline
		Ba1  & &     0            & &  0.00202(7) & &  1/4          & &  73(1) & &  1 \\
		Ba2  & &    -0.00113(14)  & & 0.33664(5)  & & 0.08965(2)  & &  87(1)  & & 1 \\
		In   & &     0            & & 0           & & 0             & &  39(1)  & & 0.968(3)\\
		Ir'  & &     0            & &  0          & &  0            & &  39    & &  0.032\\
		Ir   & &    -0.00426(7)   & &  0.33345(3) & &  0.84138(1)  & &  40(1) & &  0.897(3)\\
		In'  & &    -0.00426      & &  0.33345    & &  0.84138     & &  40    & &  0.103\\
		O1   & &     0            & &  0.4880(8)  & &  3/4          & &  38(3)  & &  1\\
		O2   & &     0.2217(10)   & & 0.2545(6)   & &  0.7433(3)    & &  38   & & 1\\
		O3   & &     0.0242(15)   & & 0.1716(5)   & &  0.9153(4)    & &  38   & & 1\\
		O4   & &     0.2272(11)   & & 0.4166(7)   & &  0.9263(3)    & &  38  & & 1\\
		O5   & &    -0.2616(11)   & & 0.4113(7)   & &  0.9079(4)    & &  38   & & 1\\
	\end{tabular}	
	\caption{Results of the refinements of the crystal structure in Ba$_3$InIr$_2$O$_9$ in
		monoclinic space group $C2/c$ with room-temperature x-ray single-crystal diffraction data.
		The lattice parameters are $a$\,=\,5.829(3)\,\AA, $b$\,=\,10.096(5)\,\AA, and $c$\,=\,14.487(7)\,\AA\ 
		for $\beta$\,=\,90$^\circ$ (see main text). 
		The parameters U$_{\rm iso}$ are given in $10^{-4}$\AA $^2$.
		Errors are given in parentheses for the last digits; if no errors are given,
		parameters are fixed by symmetry or were constrained.
		\label{tab:crystal-structure} }
\end{table}

The space group $P 6_3/m m c$ exhibits the selection rule $hh\overline{2h}l$\,$:$\,$l$\,$=$\,$2n$ 
caused by the $c$ glide plane. 
At room temperature, we observed several such forbidden reflections for which 
a $\lambda/2$ contamination was excluded by scanning with reduced generator voltage.
Typical reflections are (6 $\overline{3}$ $\overline{3}$), ($\overline{3}$ $\overline{3}$ $\overline{3}$), 
or ($\overline{6}$ 3 1).
The breaking of hexagonal symmetry can be described in the mono\-clinic space group $C2/c$ 
that was proposed for other members of the Ba$_3$$X$$M$$_2$O$_9$ family \cite{Doi04,Sakamoto06}. 
The overall structure in this family corresponds to the $6H$-BaTiO$_3$ phase, which can 
be considered as a mixing of cubic and hexagonal perovskite arrangements. 
The $X$O$_6$ octahedra form a single layer and
share a corner with neighboring $M$O$_6$ octahedra as in the cubic perovskite, see Fig.\ \ref{fig:struc}.
In contrast, the $M$ sites in the double-octahedron dimer layer share faces as in hexagonal perovskites.
In the high-symmetry arrangement in space group $P 6_3/m m c$ all oxygen ions in a plane between
two metal planes are on the same height and there are only two distinct O positions
(one on the common face of the double octahedra, one in between $X$ and $M$ ions).
Since the arrangement around the $X$ sites resembles the cubic perovskites it is not surprising
to find the main structural instability of these materials, i.e., the rotation of the $X$O$_6$ octahedra. 
This already has been proposed for several Ba$_3$$X$$M$$_2$O$_9$ compounds \cite{Doi04,Sakamoto06}
including Ba$_3$InIr$_2$O$_9$ at low temperature \cite{Dey17}. The observation of the corresponding superstructure
reflection however shows that the distortion is already present at room temperature in our single crystal.

We have refined the crystal structure in space group $C2/c$ by taking the twinning into account, see Table I.\@
The symmetry reduction from hexagonal to monoclinic symmetry results in six different twin orientations, 
whose contributions need to be summed up in the analysis. The refinement indicates a nearly equal distribution 
of these twins which results in correlations of the parameters describing the distortion. 
Furthermore the lattice constants determined on the single crystal cannot show the monoclinic distortion 
due to the twinning, so that we fix $\beta$=90$^{\circ}$.
The number of positional parameters is considerably enhanced from 7 in $P 6_3/m m c$ to 20 in $C2/c$. 
We have used the low-temperature structure \cite{Dey17} as starting parameters
in the refinement with strong damping. We find a considerable improvement of the fit when applying the lower symmetry: $wR_{all}$ ($R_{all}$) decreases from 8.41\,\%
(5.96\%) to 5.01\% (5.13\%) and the goodness of fit parameter is reduced from 5.39 to 1.87.
There is thus no doubt that the single crystal of Ba$_3$InIr$_2$O$_9$ exhibits the structural distortion 
as described in space group $C2/c$ already at room temperature.
The main element of this distortion consists of the rotation of the
InO$_6$ octahedron: the In-O-Ir bond angle amounts to 
$178.2^{\circ}$ in the hexagonal refinement and to 
$171.1^{\circ}$, $169.8^{\circ}$, and $172.2^{\circ}$ in the distorted structure. 
The structural phase transition involves the condensation of a $\Gamma_6^+$ phonon mode \cite{Stokes88} associated 
with the rotation of the octahedra around an axis parallel to the layers. Due to the degeneracy of this mode 
in the hexagonal high-temperature space group several low-temperature symmetries are possible. In $C2/c$ the octahedra rotate around a [210] direction of the hexagonal lattice, which corresponds to the monoclinic $b$ axis.
There is also some distortion in the InO$_6$ octahedron with the O-In-O bond angle deviating by up to 5$^{\circ}$
from 90$^{\circ}$, which is also frequently observed in cubic perovskites \cite{Cwik2003}.
Furthermore, there is some buckling in the O layers arising from the octahedron rotation.
However there is little impact on the bond distances.
For the intradimer Ir-Ir distance we find 2.648\,\AA\ at room temperature.

\subsection{RIXS spectra}

\begin{figure}[tb]
	\centering
	\includegraphics[width=\columnwidth]{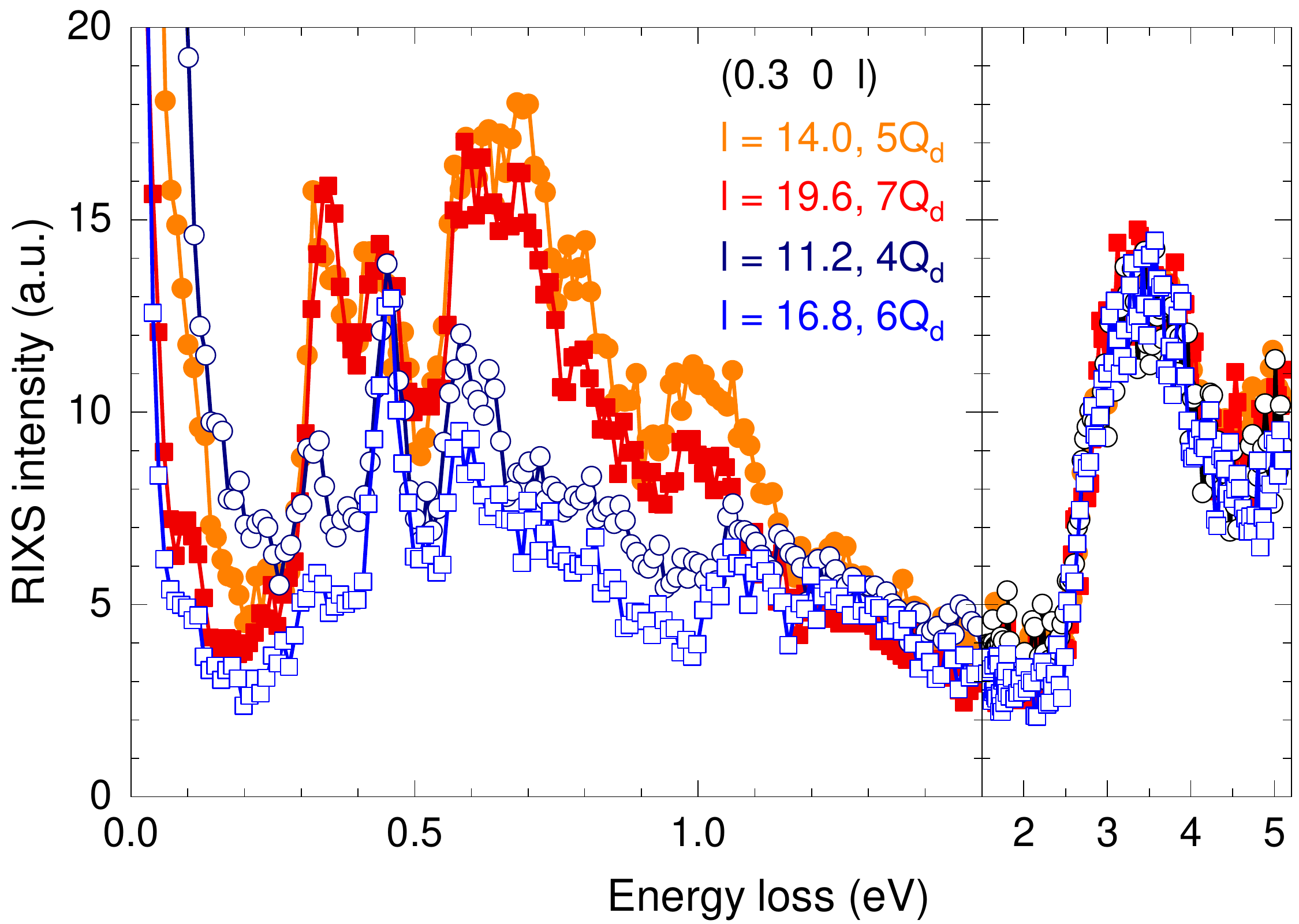}
	\caption{High-resolution RIXS spectra of Ba$_3$InIr$_2$O$_9$ at 10\,K 
	for $\mathbf{q}$\,=\,(0.3\,\,0\,\,$2.8n$)\,r.l.u.\ with integer $n$ and 
	$2.8\cdot 2\pi/c \approx \pi/d$\,=\,$Q_d$, where $d$ denotes the intradimer Ir-Ir distance.
	Left: The intra-$t_{2g}$ excitations below about 1.5\,eV show a pronounced 
	even/odd behavior with respect to $n$. The intensity is modulated with the period 
	$2Q_d$\,=\,$2\pi/d$\,=\,$5.56(2) \cdot 2\pi/c$, cf.\ Fig.\ \ref{fig:rixslscan}.
	We use a finite $h$\,=\,0.3 to avoid fulfilling the Bragg condition and the concomitant
	enhancement of the elastic line.
	Right: Excitations involving $e_g^\sigma$ orbitals are observed above 2.5\,eV.\@
	}
	\label{fig:rixst2geg}
\end{figure}

High-resolution RIXS spectra of Ba$_3$InIr$_2$O$_9$ are shown in Fig.\ \ref{fig:rixst2geg} for 
selected values of the transferred momentum $\mathbf{q}$. 
Note that we use reciprocal lattice units (r.l.u.) for indices $(h,k,l)$ 
while reciprocal space vectors are given in absolute units.
The RIXS data are analyzed in the hexagonal lattice with $P6_3/mmc$ symmetry, 
as discussed in Sec.\ \ref{sec:exp}. 
Iridates are known to show a large cubic crystal-field splitting 10\,Dq\,$\approx$\,3\,eV
between $t_{2g}$ and $e_g^\sigma$ orbitals due to the spatially extended character of the $5d$ states.
Accordingly, the RIXS features observed below 1.5\,eV can be attributed to intra-$t_{2g}$ excitations.
Excitations to $e_g^\sigma$ orbitals set in at about 2.5\,eV, see right panel of Fig.\ \ref{fig:rixst2geg},
and the increase above 4.5\,eV can tentatively be identified with charge-transfer excitations
between Ir and O sites.
Around zero energy loss, the data are dominated by the elastic line. For $\pi$ polarization of the
incident photons, the elastic line can be suppressed by choosing a scattering angle of $2\theta$\,=\,$90^\circ$.
Figure \ref{fig:rixst2geg} shows data for $\mathbf{q}$\,=\,(0.3\,\,0\,\,$l$)\,r.l.u.\
that were measured with $2\theta$\,=\,$51^\circ$, $65^\circ$, $80^\circ$, and $97^\circ$, and the 
strongest elastic line is observed for $l$\,=\,11.2 with $2\theta$\,=\,$51^\circ$.

\begin{figure}[tb]
	\centering
	\includegraphics[width=\columnwidth]{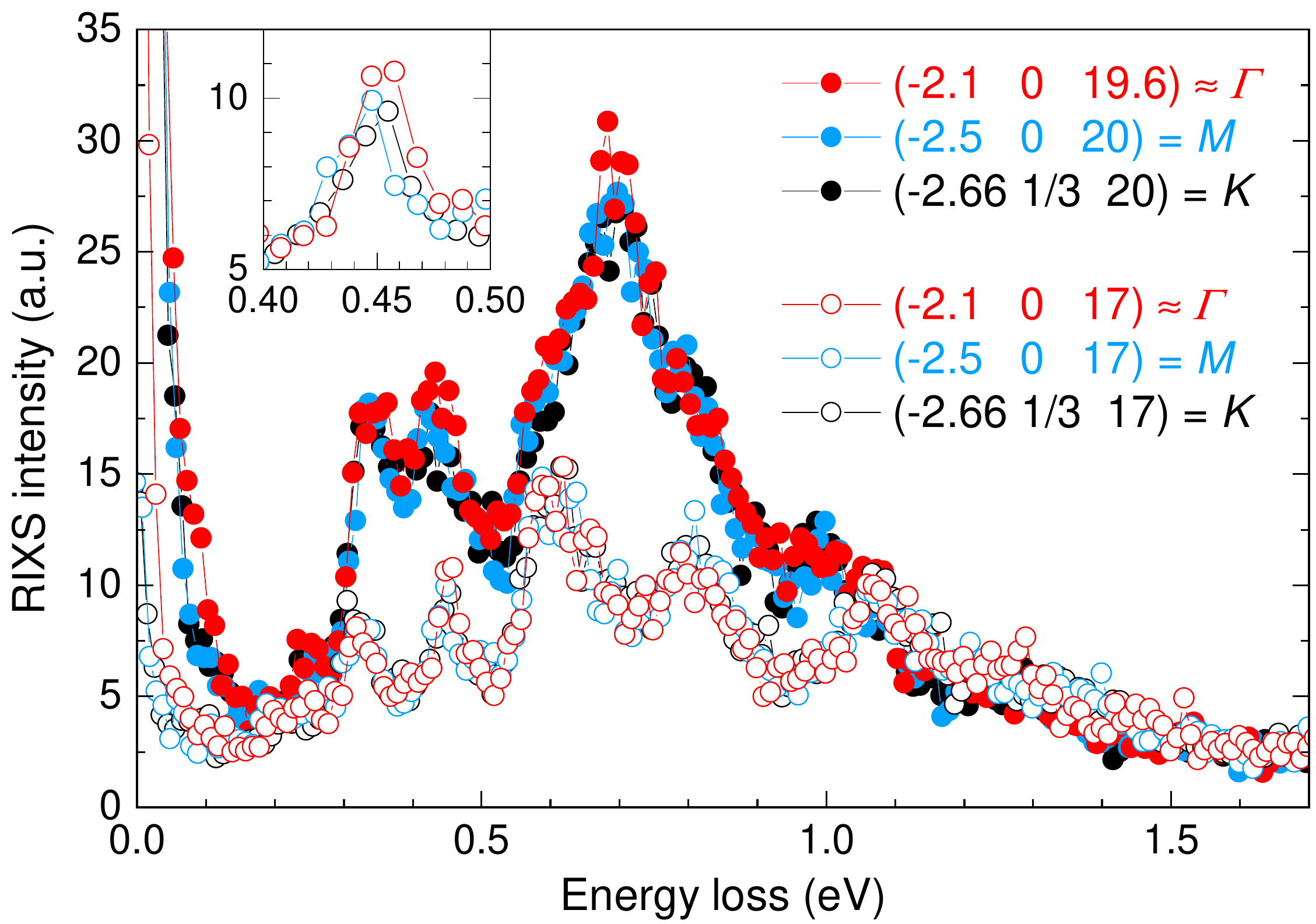}
	\caption{RIXS spectra of Ba$_3$InIr$_2$O$_9$ at the high-symmetry points $\it{\Gamma}$, $M$, 
	and $K$ for two different values of $q_c$, the component of the transferred momentum 
	parallel to the $c$ axis. For $\it{\Gamma}$, the data were measured with a small offset 
	in $\mathbf{q}$ to avoid strong elastic scattering. 
	We find a striking insensitivity to the in-plane components of the transferred momentum.
	Within the experimental energy resolution, the RIXS features do not show any considerable
	dispersion. The largest change of up to 10\,meV is found for the peak at 0.45\,eV (see inset).
	The data are dominated by the strong dependence of the intensity on $q_c$,
	very similar to the behavior shown in Fig.\ \ref{fig:rixst2geg}. 
	The integer values $l$\,=\,17 and 20 were chosen since they are close 
	to extrema of the intensity modulation at $6Q_d$ and $7Q_d$, see Fig.\ \ref{fig:rixslscan}a. 
	Compared to Fig.\ \ref{fig:rixst2geg}, the larger overall intensity stems from the 
	difference in $h$, see Fig.\ \ref{fig:rixslscan}b. 
	}
	\label{fig:rixsGMK}
\end{figure}

The RIXS spectrum of intra-$t_{2g}$ excitations is very rich in Ba$_3$InIr$_2$O$_9$.
Figure \ref{fig:rixst2geg} shows the effect of changing $q_c$, the component of the transferred
momentum parallel to the $c$ axis, and Fig.\ \ref{fig:rixsGMK} addresses a possible dispersion
within the hexagonal plane, perpendicular to $c$, showing data for different 
high-symmetry points for two different values of $q_c$. 
We find that the intensity of the intra-$t_{2g}$ excitations strongly varies with $q_c$,
while the peak energies do not depend considerably on $\mathbf{q}$.
For constant $q_c$, the data basically fall on top of each other in Fig.\ \ref{fig:rixsGMK}.
The largest effect of $\mathbf{q}$ on the peak energy is observed for the feature at about
0.45\,eV, which sets an upper limit of 10\,meV for a possible dispersion,
see the inset of Fig.\ \ref{fig:rixsGMK}.
This reflects the fact that interdimer hopping of the excited states is small.
Since such interdimer hopping interactions form the microscopic basis for magnetic 
exchange interactions, a dispersion smaller than 10\,meV suggests that also exchange 
interactions between dimers are small. This agrees with the small Curie-Weiss temperature 
$\Theta_{\rm CW}$\,=\,$-6.8$\,K derived from an analysis of magnetic susceptibility data 
and the small temperature scale of about 2\,K for the occurrence of correlated 
spin-liquid-like behavior \cite{Dey17}, which suggests interdimer exchange interactions 
of less than 1\,meV.

\subsection{RIXS interference patterns and quasimolecular orbitals}
\label{sec:interference}

\begin{figure}[tb]
	\centering
	\includegraphics[width=0.9\columnwidth]{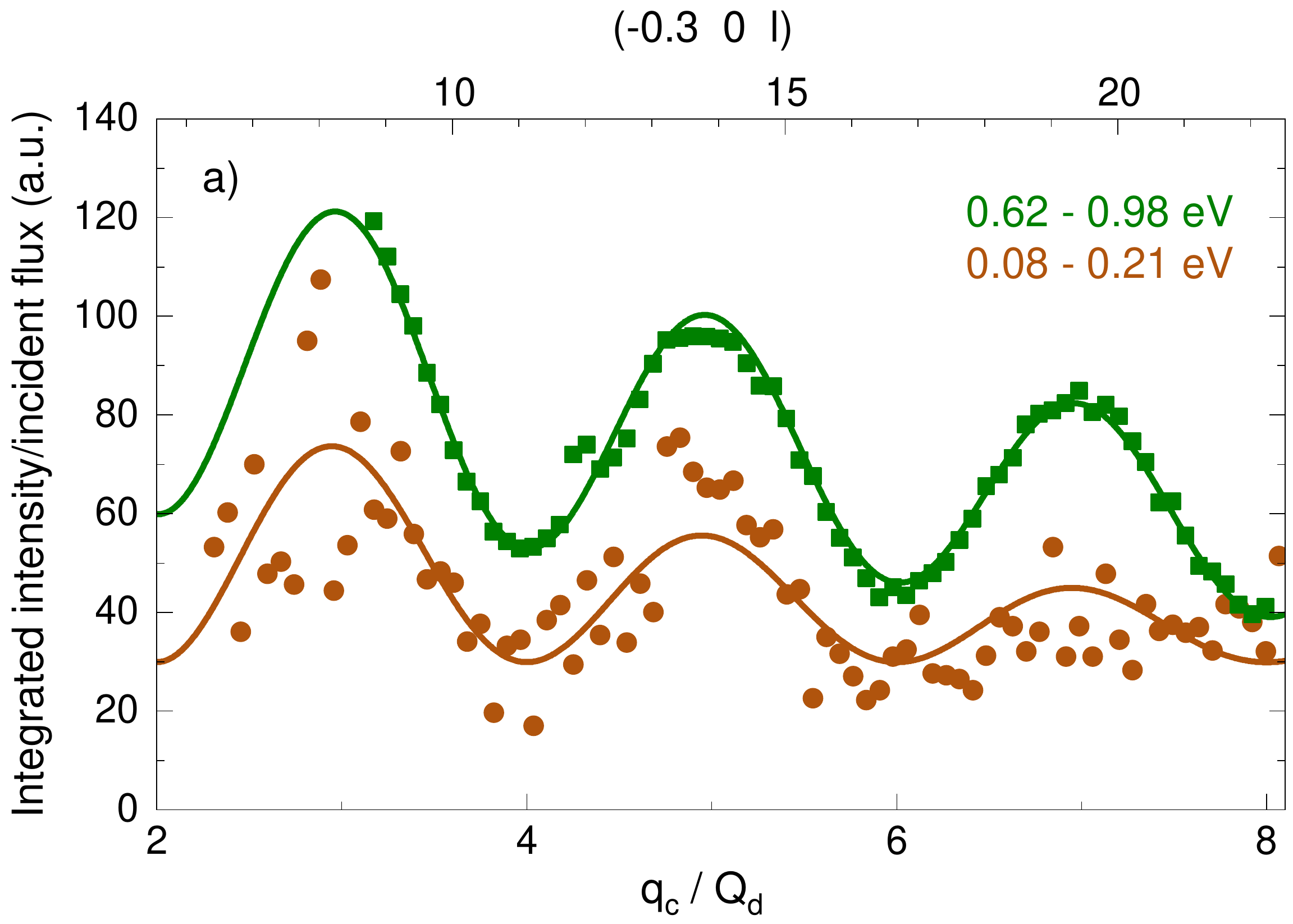}
	\includegraphics[width=0.9\columnwidth]{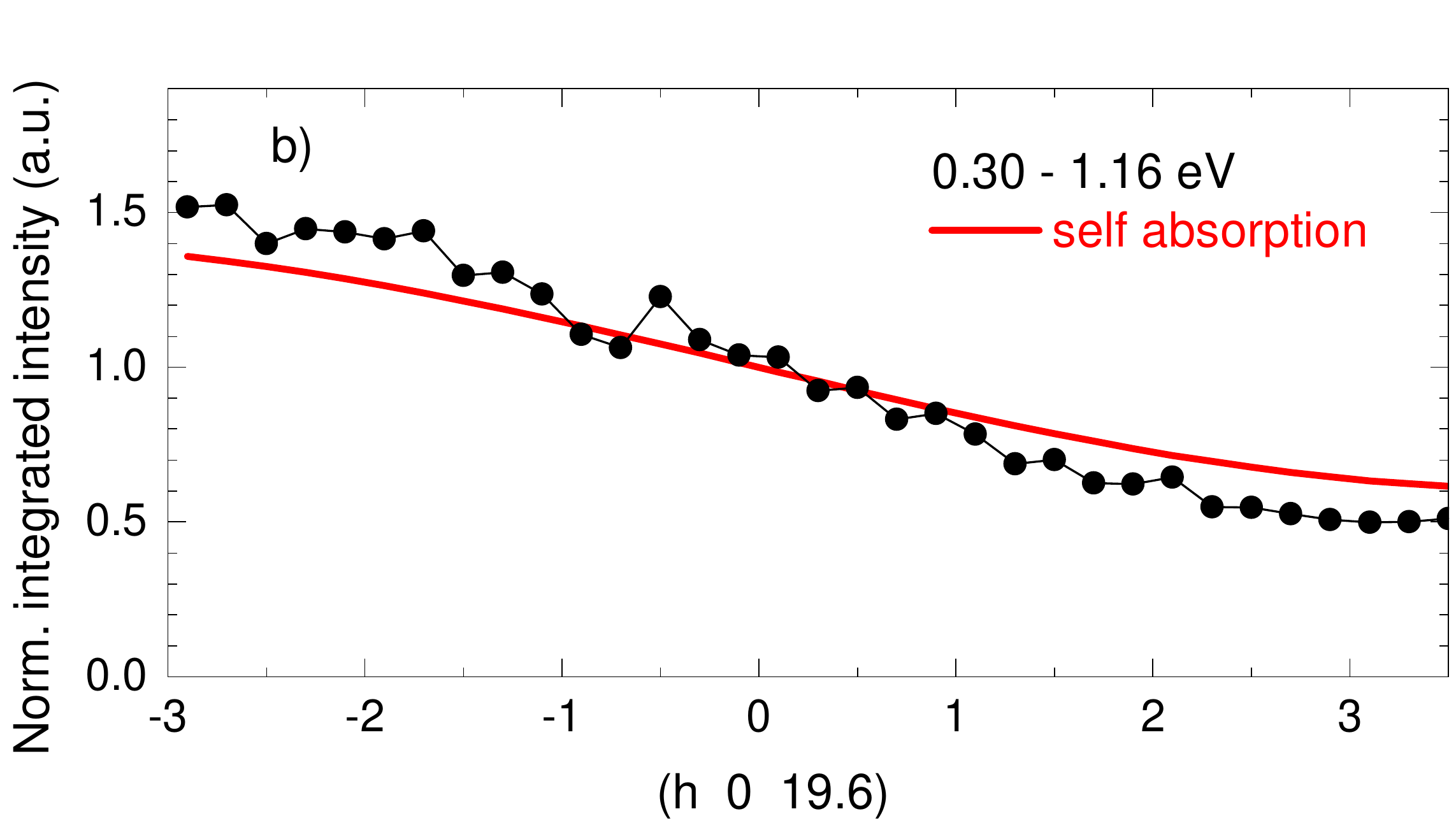}
	\caption{RIXS intensity $I(\mathbf{q})$ of Ba$_3$InIr$_2$O$_9$ at 10\,K.\@
		a) Data for $\mathbf{q}$\,=\,(-0.3\,\,0\,\,$l$)\,r.l.u.\  (top axis) 
		reveal a sinusoidal intensity modulation that unambiguously demonstrates 
		the quasimolecular dimer character of the electronic structure. 
		The bottom axis emphasizes the period $2Q_d$\,=\,$2\pi/d$\,=\,$5.56 \cdot 2\pi/c$ 
		which is incommensurate with the Brillouin zone. 
		Solid lines show damped sinusoidal fits, see Eq.\ (\ref{eq:Iqc}).
		Green symbols: Low-resolution data, $\Delta E$\,=\,0.36\,eV,
		covering the range 0.62 -- 0.98\,eV.\@ Integration time was 10\,s per $\mathbf{q}$ point.
		Brown symbols: Data integrated from 0.08 to 0.21\,eV with the high-resolution setup,
		$\Delta E$\,=\,25\,meV, and 20\,s integration time per $\mathbf{q}$ point.
		The advantage of the low-resolution setup is the enhanced signal-to-noise ratio.
		Data were normalized by the incident flux.
		b) As a function of $h$, the RIXS intensity, integrated from 0.3 to 1.16\,eV 
		shows monotonic behavior. Data were normalized to the value at $h$\,=\,0.
		The solid red line depicts the expected behavior due to self-absorption.
	}
	\label{fig:rixslscan}
\end{figure}

The absence of strong dispersion shows that interdimer hopping is small. In the following,
we address the role of intradimer hopping for the electronic structure.
RIXS provides an unambiguous tool to distinguish different limits.
For dominant $U$ and small intradimer hopping $t$, the presence of local $j$\,=\,$1/2$ or
$j$\,=\,$0$ moments can be detected via characteristic narrow features in the RIXS spectra \cite{Gretarsson13,Rossi17,Liu12,Revelli19b,Reig20,Khan21,Moretti14,Yuan17,Kusch18,Nag18}.
For the opposite limit of dominant intradimer hopping, the observation of sinusoidal 
double-slit-type RIXS interference patterns provides a litmus test for a quasimolecular 
electronic structure \cite{Revelli19}.
As described in the following, the pronounced modulation of the RIXS 
intensity $I(q_c)$ depicted in Figs.\ \ref{fig:rixst2geg}, \ref{fig:rixsGMK}, and 
\ref{fig:rixslscan}a is caused by such a double-slit-type interference and demonstrates 
the quasimolecular dimer character. 
To explain the occurrence of such interference patterns, we address the RIXS process at the
Ir $L_3$ edge \cite{Ament11}, in which an incident x-ray photon excites an electron out of a
strongly localized $2p$ core level into an empty $5d$ level. In particular, the choice of
11.215\,keV for the incident energy resonantly enhances absorption into a $t_{2g}$ orbital 
in the $5d$ shell.  This intermediate state decays by emitting a photon with lower energy,
and the $2p$ core hole is filled by a $5d$ valence electron. The resonance enhancement particularly
boosts the sensitivity to intra-t$_{2g}$ excitations, which may show orbital or magnetic character.
In other words, resonant inelastic scattering proceeds from the ground state via an intermediate state
with a strongly localized $2p$ core hole to a final state with an excitation that may be delocalized,
for instance a magnon.
For a given excitation, the total RIXS amplitude is a coherent sum running over the scattering
processes on all Ir sites over which this final excited state is delocalized \cite{Gelmukhanov94,Ma94}.
If the excited state is a quasimolecular orbital excitation in which electrons are 
delocalized over two Ir sites, the interference between x-ray photons emitted from these 
dimer sites yields a sinusoidal interference pattern as a function of the transferred 
momentum {\bf q}, in close analogy to the case of a double-slit experiment \cite{Ma95}.
The same applies to, e.g., magnetic excitations if spin correlations are restricted 
to two adjacent sites, and the corresponding sinusoidal interference pattern 
was observed in RIXS on the honeycomb Kitaev materials Na$_2$IrO$_3$ and 
$\alpha$-Li$_2$IrO$_3$ \cite{Revelli20}.

This explains the strong dependence of the intensity on $\mathbf{q}\cdot\mathbf{d} = q_c\,d$, 
cf.\ Fig.\ \ref{fig:rixst2geg}, where $\mathbf{d}$\,=\,(0,0,$d$) is the vector connecting 
two Ir sites within a dimer. The sinusoidal character of the intensity modulation,
\begin{equation}
\label{eq:Iqc}
I(q_c) \propto \sin^2(q_c \,d/2) + {\rm const}\, ,
\end{equation}
is revealed by measurements of the RIXS intensity $I(q_c)$ for fixed energy loss,
see Fig.\ \ref{fig:rixslscan}a. The observed period 
$2Q_d$\,=\,$2\pi/d$\,=\,$5.56(2) \cdot 2\pi/c$
yields an Ir-Ir distance $d$\,=\,2.601(9)\,\AA\ at 10\,K,
in excellent agreement with the neutron diffraction result 2.599(4)\,\AA\ at 3.4\,K \cite{Dey17}.
The sinusoidal form of the modulation and the value of the period $2Q_d$ unambiguously prove
that the intra-$t_{2g}$ excitations are delocalized over the two sites of an Ir dimer,
hence the electronic structure of Ba$_3$InIr$_2$O$_9$ needs to be described in a quasimolecular picture.

The $\sin^2(q_c \,d/2)$ interference pattern deviates from the $\cos^2(q_c\,d/2)$ behavior
well known from a conventional elastic double-slit experiment. For RIXS on dimers,
the observation of a $\cos^2(q_c\,d/2)$ or $\sin^2(q_c\,d/2)$ interference pattern
contains information on the symmetry of the investigated states \cite{Revelli19}.
This can be explained by the dipole selection rules for both photon absorption and photon emission.
Consider, e.g., an even ground state.
In the  limit $\mathbf{q}$\,=\,0, RIXS excitations are allowed via an odd intermediate state to an
even final state. For finite $\mathbf{q}$, we additionally have to consider the geometrical path difference
for scattering events on the two adjacent Ir sites. Together, this yields a
$\cos^2(q_c\,d/2)$ ($\sin^2(q_c\,d/2)$) interference pattern for even (odd) final states
in the case of an even ground state.

The Ir$_2$O$_9$ dimers show mirror symmetry but no inversion symmetry, 
as the two face-sharing octahedra are rotated with respect to each other by $\pi$ around $c$.
Parity therefore is not a good quantum number of the eigenstates. This results in interference patterns
with mixed behavior, $I(q_c) \propto u \sin^2(q_cd/2) + v \cos^2(q_cd/2)$, cf.\ Eq.\ (\ref{eq:Iqc}).
In RIXS on Ba$_3$InIr$_2$O$_9$, the $\sin^2(q_c\,d/2)$ behavior strongly prevails for the intra-$t_{2g}$ 
excitations below about 1\,eV, similar to the case of Ba$_3$CeIr$_2$O$_9$ \cite{Revelli19}.
Figure \ref{fig:rixslscan}a demonstrates this dominant $\sin^2(\pi\,q_c/2Q_d)$ behavior for 
the integrated intensity plotted as a function of $q_c/Q_d$, 
and the spectra depicted in Figs.\ \ref{fig:rixst2geg} and \ref{fig:rixsGMK} 
corroborate this behavior for energies below about 1.1\,eV.\@
These spectra were recorded for values of $q_c \approx nQ_d$ that correspond to extrema 
of the interference pattern.
RIXS spectra for, e.g., odd $n$ agree with each other over almost the entire range of 
energy loss. Below about 1.1\,eV, the intensity is higher for odd $n$ and the data 
strongly differ from the spectra for even $n$. 
Even though parity is not a good quantum number, we find that excitations from 
bonding to antibonding orbitals, as defined in Sec.\ \ref{sec:discuss}, predominantly show 
$\sin^2(q_c\,d/2)$ behavior. The experimental result hence agrees with a $j_{\rm dim}$\,=\,3/2 
ground state as depicted in Fig.\ \ref{fig:sketch}, where all three holes occupy bonding orbitals 
and most of the excitations correspond to transitions to antibonding ones. 
This conclusion is supported by the analysis in Sec.\ \ref{sec:discuss}.

Wang \textit{et al.} \cite{Wang18} reported on RIXS data of Ba$_5$AlIr$_2$O$_{11}$
which also shows face-sharing IrO$_6$ octahedra that form Ir dimers with three holes.
However, these Ir dimer sites are crystallographically nonequivalent,
resulting in charge disproportionation \cite{Streltsov17}. The formation of
quasimolecular orbitals, predicted by Streltsov, Cao, and Khomskii \cite{Streltsov17},
has been derived from the RIXS spectra by arguing that they deviate from those of 
local moments and by comparing with cluster calculations.
Data addressing the RIXS intensity as a function of $\mathbf{q}\cdot \mathbf{d}$ 
were not reported. Measured with an energy resolution 
of 80\,meV, the RIXS spectra of Ba$_5$AlIr$_2$O$_{11}$ show five broad peaks at 
0.18, 0.32, 0.57, 0.74, and 1.13\,eV.\@ They resemble our results obtained at an 
interference minimum assuming a somewhat reduced energy scale.

\subsection{In-plane momentum dependence}

\begin{figure}[tb]
	\centering
	\includegraphics[width=\columnwidth]{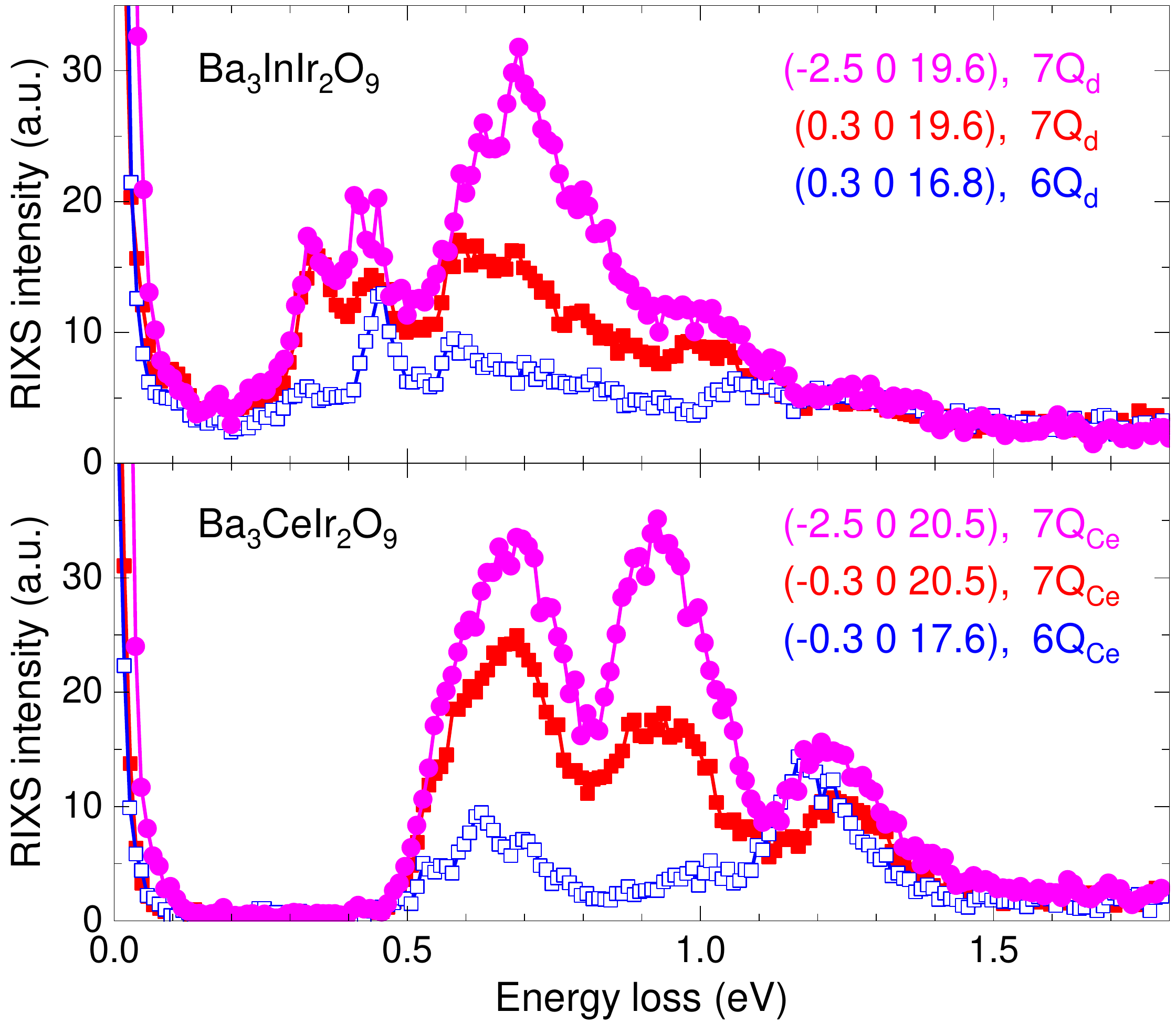}
	\caption{Comparison of RIXS spectra of Ba$_3$InIr$_2$O$_9$ (top) and Ba$_3$CeIr$_2$O$_9$ (bottom).
		The Ce compound has two holes per dimer, i.e., no partially filled quasimolecular orbitals,
		and exhibits a clear onset of RIXS intensity at about 0.5\,eV.\@
		The different Ir-Ir distance $d_{\rm Ce}$\,=\,$c_{\rm Ce}/5.83$ gives rise to a slightly different
		period $2Q_{\rm Ce}$ of the interference pattern.
	}
	\label{fig:rixsInCe}
\end{figure}

In contrast to the pronounced sinusoidal dependence of $I(q_c)$, the RIXS intensity shows a moderate,
monotonic intensity change as a function of $h$, see Fig.\ \ref{fig:rixslscan}b.
Any change of $\mathbf{q}$ for fixed incident energy requires to change the experimental geometry.
This affects matrix elements and self-absorption effects.
The latter can be described by \cite{Minola15} $1/(1+\sin\theta_{\rm in}/\sin\theta_{\rm out})$,
where $\theta_{\rm in}$ and $\theta_{\rm out}$ denote the angle between the sample surface and the
incident and emitted beams, respectively. For grazing incidence, i.e., small $\theta_{\rm in}$,
the incident photons are absorbed close to the surface. This facilitates the escape of the re-emitted photons
and results in an enhanced signal. The overall $h$ dependence of the intensity agrees with this expectation,
as depicted by the red line in Fig.\ \ref{fig:rixslscan}b.

\begin{figure}[tb]
	\centering
\includegraphics[width=\columnwidth]{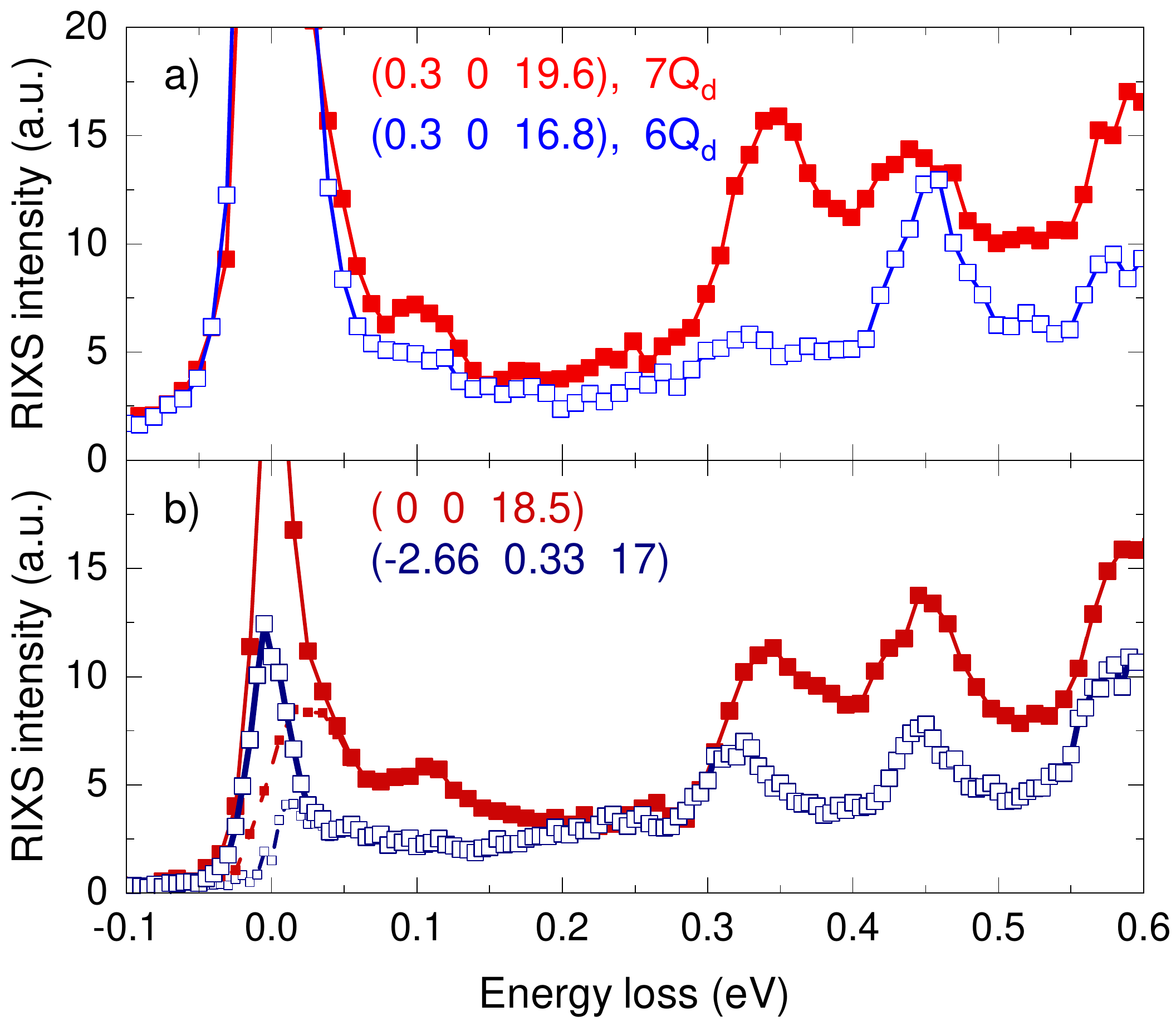}
	\caption{Low-energy RIXS spectra of Ba$_3$InIr$_2$O$_9$ at 10\,K.\@
		a) Zoom in on two data sets shown in Fig.\ \ref{fig:rixst2geg}, measured with 10\,meV step size
		for $q_c$ values on extrema of the interference pattern, see Fig.\ \ref{fig:rixslscan}.
		Integration time was 20\,s per detector image.
	b) Data collected with $2\theta$\,=\,$90^\circ$ to suppress the elastic line. 
	With the given incident energy, this fixes $|\mathbf{q}|$. With this restriction, 
	we can reach $q_c$\,=\,$18.5 \cdot 2\pi/c \! \approx \! 6.7\,Q_d$, reasonably close 
	to the interference maximum at $7Q_d$, and 
	\mbox{$\mathbf{q}$\,=\,(-2.66\,\,1/3\,\,17)\,r.l.u}.\ with $q_c$\,$\approx$\,$6Q_d$, 
	an interference minimum. 
	For a quantitative comparison, one has to take into account that the intensity at $6.7Q_d$
	(dark red symbols) is roughly 10\,\% smaller than at $7Q_d$, see Fig.\ \ref{fig:rixslscan}a.
	Self-absorption has been corrected.
	Dashed lines with small symbols show an estimate of the inelastic contribution obtained 
	by subtracting the already suppressed elastic line.
	Dark blue (dark red): 5\,meV (10\,meV) step size and 40\,s (30\,s) integration time.
	}
	\label{fig:rixs3}
\end{figure}

Additional matrix-element effects are revealed by comparing spectra for different $h$,
see top panel of Fig.~\ref{fig:rixsInCe}. For the same value of $l$, the intensity is larger for
$h$\,=\,$-2.5$ than for $h$\,=\,$0.3$, but this effect is most pronounced around 0.7\,eV.
Remarkably, a similar behavior is observed from 0.5 to 1.1\,eV in the sister compound
Ba$_3$CeIr$_2$O$_9$ \cite{Revelli19} with two holes per dimer, see bottom panel of Fig.\ \ref{fig:rixsInCe}.
In Ba$_3$CeIr$_2$O$_9$, the two prominent peaks at about 0.7\,eV and 0.9\,eV show $\sin^2(q_cd/2)$
behavior \cite{Revelli19}, very similar to Ba$_3$InIr$_2$O$_9$. In reasonable approximation, 
these two peaks of Ba$_3$CeIr$_2$O$_9$ can be described in a single-particle picture as excitations 
from the bonding quasimolecular orbital formed from local $j$\,=\,1/2 states
to bonding quasimolecular orbitals built from local $j$\,=\,$3/2$ states \cite{Revelli19}.
For the 0.7\,eV peak in Ba$_3$InIr$_2$O$_9$, the similar energy range and the similar dependence 
on $h$ suggest a related character. Our analysis supports a contribution of such states 
but finds also further excitations in this energy range, see Sec.\ \ref{sec:discuss}.

\subsection{Low-energy behavior}
\label{sec:lowE}

In Ba$_3$CeIr$_2$O$_9$, the RIXS intensity vanishes below 0.5\,eV, which in a single-particle 
picture reflects the energy difference between filled and empty quasimolecular 
orbitals \cite{Revelli19}.
In contrast, Ba$_3$InIr$_2$O$_9$ exhibits finite RIXS intensity also at low energy, see Fig.\ \ref{fig:rixs3}.
Previously, the existence of excitations at about 10\,meV and 40\,meV has been estimated from
an analysis of the magnetic susceptibility \cite{Dey17}.
To better resolve the low-energy features, we suppressed the elastic line by choosing a scattering angle
$2\theta$\,=\,$90^\circ$ with incident $\pi$ polarization, see Fig.\ \ref{fig:rixs3}b.
Dashed lines with small symbols estimate the inelastic contribution after subtraction of the
remaining elastic line.
With the given energy resolution of 25\,meV, the data in Fig.\ \ref{fig:rixs3}b show that the RIXS intensity
remains finite for any energy loss below 0.3\,eV.\@
The data for $l$\,=\,17 appear continuum-like but the result for $l$\,=\,18.5 points toward peaks
at about 0.03\,eV, 0.10\,eV, and 0.23\,eV.\@

\begin{figure}[t]
	\centering
	\includegraphics[width=\columnwidth]{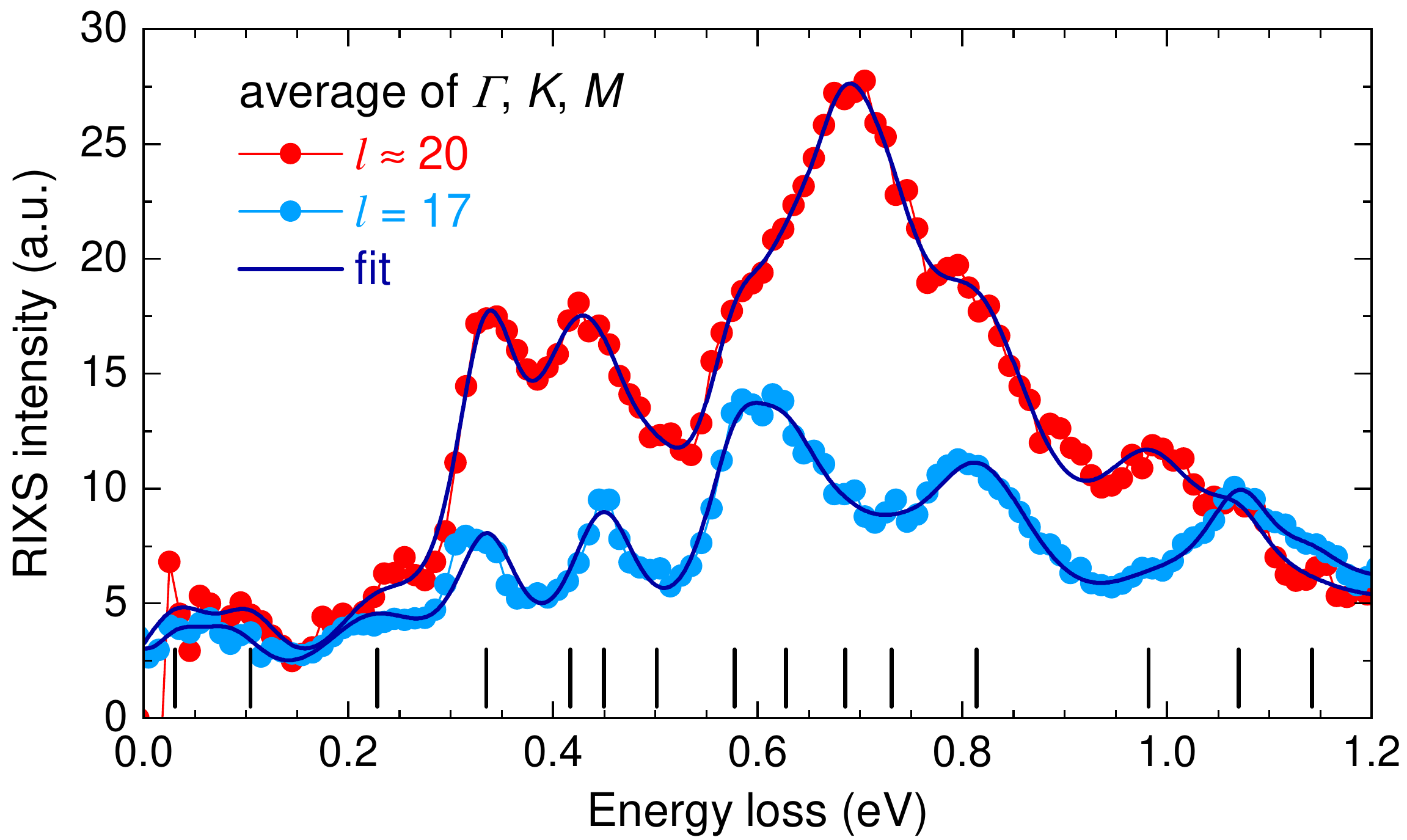}
	\caption{Oscillator fit (solid lines) of RIXS data averaged over the spectra measured 
	at about $\it{\Gamma}$, $M$, and $K$, cf.\ Fig.~\ref{fig:rixsGMK}. We distinguish 
	$l \approx 20$ and $17$ 
	which roughly correspond to the extrema of the interference pattern at 
	$q_c$\,=\,$7Q_d$ and $6Q_d$, cf.\ Fig.\ \ref{fig:rixslscan}a. 
	The elastic line has been subtracted. 
	Parameters are given in Table \ref{tab:fit}. 
	Vertical bars denote the excitation energies found in the fit.
	}
	\label{fig:rixsdiff}
\end{figure}

Within a dimer, low-energy excitations occur in the vicinity of the phase transition from 
$j_{\rm dim}$\,=\,1/2 to $j_{\rm dim}$\,=\,3/2, see Sec.\ \ref{sec:discuss}.
The peak energies and the RIXS resolution are much larger than the interdimer magnetic 
energy scale given by the peak in the specific heat at about 2\,K \cite{Dey17}. 
Therefore a direct connection between the observed low-energy excitations and the magnetic 
fluctuations of a possible spin-liquid ground state can be ruled out. 
However, it is tempting to speculate that 
a small but finite magnetic contribution to the low-energy RIXS intensity 
may arise due to In/Ir site mixing. Sakamoto \textit{et al.} \cite{Sakamoto06} studied 
powder samples of Ba$_3$$X$Ir$_2$O$_9$ for different metals $X$ including In and found 
that $X$-Ir cation disorder is only relevant for $X$\,=\,Ti$^{4+}$ and Sc$^{3+}$.
This can be explained by the similar ionic radii of Ir$^{4+}$, Ti$^{4+}$, and Sc$^{3+}$
\cite{Sakamoto06}. Larger $X$ ions such as Ce$^{4+}$ or In$^{3+}$ as well as the different 
valence of In and Ir suppress cation disorder.
For Ba$_3$InIr$_2$O$_9$, Dey \textit{et al.} \cite{Dey17} discussed that the structural data of 
powder samples are well described by fully ordered structure models without cation disorder, 
while deliberately adding site mixing yields an estimate of 2.8(5)\,\%.
Our single-crystal x-ray diffraction data indicate about 1-2\,\% of Ir ions on the In $2a$ site 
\textit{between} dimers, see Sec.\ \ref{sec:struc}. 
Even though the estimated density of such Ir moments is small, their contribution 
may be relevant since each of them is coupled to six neighboring dimers. 
The $2a$ sites are connected with the $4f$ dimer sites in a corner-sharing configuration
with close to $180^\circ$ Ir-O-Ir bonds, and Ir moments on the $2a$ sites are expected 
to enhance interdimer coupling. The bonding geometry is closely related to the case of 
Sr$_2$IrO$_4$ which shows large exchange interactions and magnetic excitations up to 
about 0.2\,eV \cite{Kim12,Kim14}. 
In Ba$_3$InIr$_2$O$_9$, the sinusoidal modulation observed for integration from 0.08 to 0.21\,eV 
demonstrates the existence of dimer excitations at low energy, see Fig.\ \ref{fig:rixslscan}a. 
However, the $l$ dependence is small between 0.2\,eV and 0.3\,eV, see Fig.\ \ref{fig:rixs3}b. 
A small magnetic contribution caused by In-Ir site disorder hence cannot be ruled out.

\section{Discussion}
\label{sec:discuss}

\begin{table}[t]
	\begin{tabular}[t]{cp{2mm}cp{2mm}cp{2mm}cp{2mm}cp{2mm}c}
		energy [eV] && $I_{6Q_d}$ && $I_{7Q_d}$ && $\frac{I_{7Q_d}-I_{6Q_d}}{I_{7Q_d}+I_{6Q_d}}$ && width [meV] && $\mu$ \\ \hline
		0.03   &&      &&      &&       &&     &&  \\
		0.10   && 2.4  && 3.1  &&  0.14 &&  81 && 2.0  \\
		0.23   && 3.7  && 3.9  &&  0.02 && 106 && 4.8  \\
		0.34   && 6.8  && 13.9 &&  0.34 &&  76 && 2.0 \\
		0.42   && 0.3  && 10.7 &&  0.94 &&  99 && 1.6  \\
		0.45   && 7.2  && 5.1  && -0.17 &&  82 && 1.9 \\
		0.50   && 0.5  && 4.7  &&  0.79 &&  90 && 1.9 \\
		0.58   && 7.9  && 9.4  &&  0.09 &&  80 && 1.5 \\
		0.63   && 7.6  && 8.5  &&  0.06 &&  93 && 1.9 \\
		0.69   && 3.0  && 16.8 &&  0.70 &&  97 && 1.8 \\
		0.73   && 1.4  && 6.6  &&  0.66 &&  83 && 1.6 \\
		0.81   && 9.1  && 14.2 &&  0.22 && 148 && 1.8 \\
		0.98   && 2.1  && 7.4  &&  0.56 && 137 && 1.3 \\
	\end{tabular}	
	\caption{Parameters of the fit depicted in Fig.\ \ref{fig:rixsdiff}, where only the 
	peak intensities are allowed to vary with $q_c$. 		
	The Pearson VII oscillator describes a Lorentzian (Gaussian) line shape for 
	$\mu$\,=\,1 ($\infty$) \cite{Wang05Pearson}.
	Above about 1\,eV it is not possible to distinguish different features.
	The lowest excitation energy of 0.03\,eV is estimated from the data in Fig.\ \ref{fig:rixs3}b.
\label{tab:fit} }
\end{table}

Having established the quasimolecular nature of the electronic states, we aim to achieve a 
microscopic understanding of the electronic structure of Ba$_3$InIr$_2$O$_9$. To this end, 
we compare our RIXS data with exact diagonalization results. Typical for iridates and other 
$5d$ compounds, the electronic structure is governed by the interplay of Coulomb interactions, 
spin-orbit coupling, hopping interactions, and the non-cubic part of the crystal field.
However, for the cluster Mott insulator Ba$_3$InIr$_2$O$_9$ the focus is in particular on the 
competition between hopping and spin-orbit coupling. Is hopping strong enough to counteract 
the effects of spin-orbit coupling?
In general, individual $5d^5$ Ir$^{4+}$ sites show local moments with $j$\,=\,1/2 and 3/2.
In the case of two holes per dimer as in Ba$_3$CeIr$_2$O$_9$ with Ir$^{4+}$, (anti-) bonding 
states of these spin-orbit entangled $j$ states provide a good starting point for the 
description \cite{Revelli19}.
Below we show that the same applies to Ba$_3$InIr$_2$O$_9$, as sketched in Fig.\ \ref{fig:sketch}, 
giving rise to the rich excitation spectrum observed in RIXS.\@

\begin{figure}[t]
	\centering
\includegraphics[width=\columnwidth]{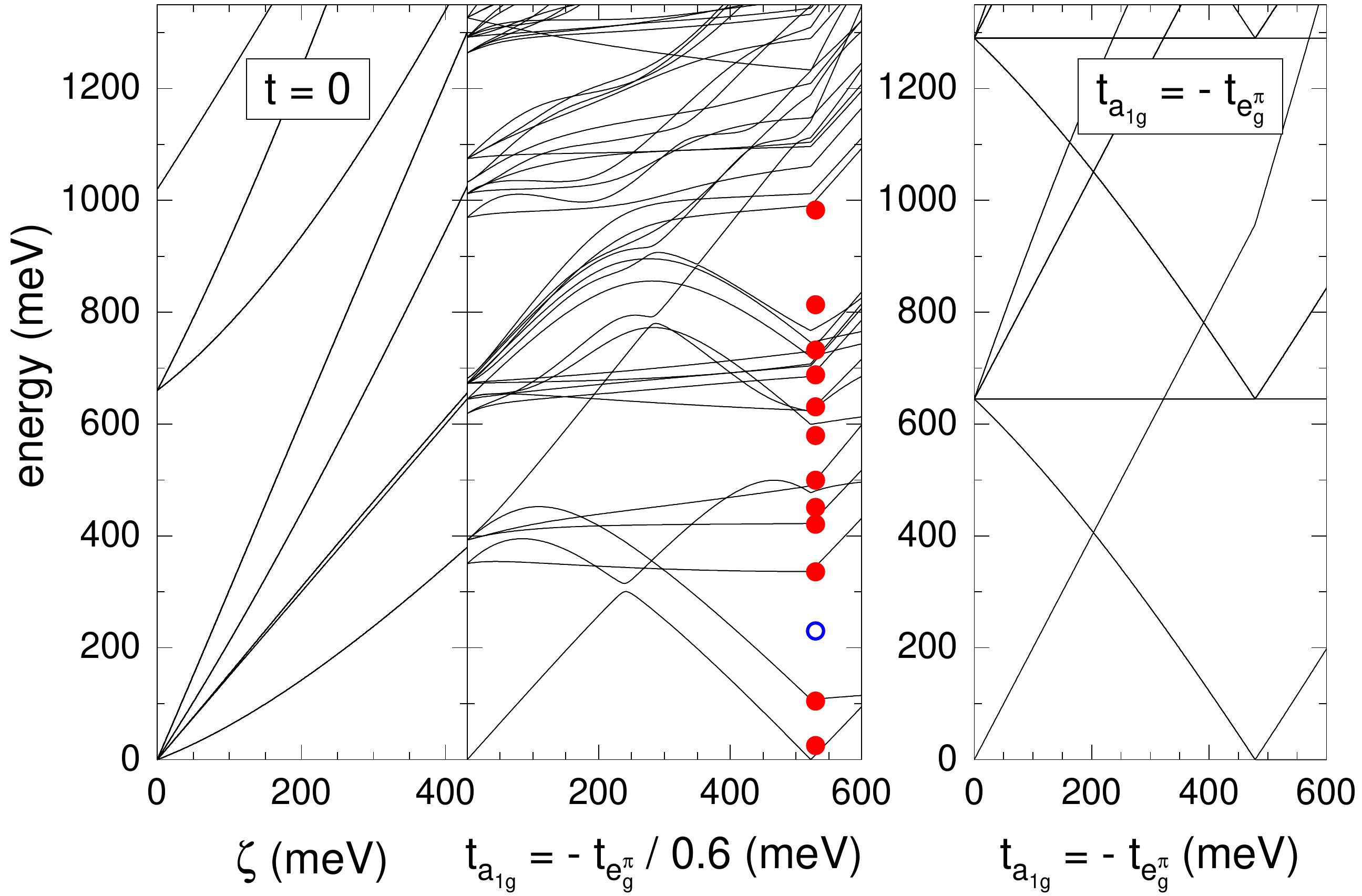}
	\caption{Excitation energies of the many-body states calculated by exact 
	diagonalization for $U$\,=\,1.5\,eV.\@
	Left: Dependence on $\zeta$ for $t$\,=\,$\Delta_{\rm CF}$\,=\,0 and $J_H$\,=\,0.33\,eV.\@ 
	For $\zeta$\,=\,0.43\,eV, the lowest excitation energies 
	of about 0.4\,eV and 0.7\,eV agree with the dominant characteristic RIXS features of 
	$t_{2g}^4$ iridates \cite{Yuan17,Kusch18,Nag18} and with the spin-orbit exciton of 
	$t_{2g}^5$ iridates at $3/2\zeta$. 
	Right: Effect of hopping in the simplified case	$t$\,=\,$t_{a_{1g}}$\,=\,$-t_{e_g^\pi}$
	for $\zeta$\,=\,0.43\,eV and $J_H$\,=\,$\Delta_{\rm CF}$\,=\,0, where hopping is 
	block-diagonal in the $j$ basis. Notably, several excitation energies are independent of $t$.
	Middle: Excitation energies as a function of hopping $t_{a_{1g}}$ for the realistic 
	parameter set with $\zeta$\,=\,0.43\,eV, $J_H$\,=\,0.33\,eV, $\Delta_{\rm CF}$\,=\,0.1\,eV,
	and $t_{e_g^\pi}/t_{a_{1g}}$=\,$-0.6$. 
	For these parameters, the ground state changes at about 0.52\,eV, and good agreement 
	with the experimental peak energies (symbols) is obtained for $t_{a_{1g}}$=\,0.53\,eV.\@ 
	The weak RIXS feature at 0.23\,eV (open blue symbol) is 
	tentatively attributed to Ir/In site disorder, see Sec.\ \ref{sec:lowE}.
	}
	\label{fig:3holeQMOs}
\end{figure}

First, we determine the peak energies by fitting the RIXS data using Pearson VII oscillators \cite{footnotePearson}.
For the fit, we considered the dispersion as negligible, as discussed above, and averaged 
the data of Fig.\ \ref{fig:rixsGMK} over $\it{\Gamma}$, $M$, and $K$ for constant $q_c$  
to improve statistics.
Motivated by the intensity modulation depicted in Fig.\ \ref{fig:rixslscan}a, we focus on
$q_c/(2\pi/c) \approx 20$ and 17, i.e., close to a maximum and a minimum of interference.
In the fit, only the peak intensities are allowed to change as a function of $q_c$,
all other parameters are treated as independent of $q_c$.
The fit describes the RIXS spectra very well, see Fig.\ \ref{fig:rixsdiff}. 
The corresponding parameters are summarized in Table \ref{tab:fit}, 
and the peak energies are plotted as symbols in the middle panel of Fig.\ \ref{fig:3holeQMOs}.

The interaction Hamiltonian with on-site Coulomb
repulsion $U$ and Hund's coupling $J_H$ reads 
\cite{Perkins14}
\begin{eqnarray}
\nonumber
H_C & = & U \, \sum_{i,\alpha} n_{i\alpha \uparrow} n_{i\alpha \downarrow}
+  \frac{1}{2} (U-3J_H) \sum_{i,\sigma,\alpha\neq\alpha^\prime} n_{i\alpha \sigma} n_{i\alpha^\prime \sigma} \\
\nonumber
& + & (U-2J_H) \sum_{i,\alpha\neq\alpha^\prime} n_{i\alpha \uparrow} n_{i\alpha^\prime \downarrow}
\\
\nonumber
& + & (U-2J_H)\sum_{i} \Big( 15-5\sum_{\alpha,\sigma} n_{i\alpha \sigma} \Big) 
\\
\nonumber
& + & J_H \sum_{\alpha \neq \alpha'} \left( c^\dagger_{\alpha \uparrow} c^\dagger_{\alpha \downarrow} c_{\alpha'\downarrow} c_{\alpha'\uparrow}
- c^\dagger_{\alpha\uparrow} c_{\alpha\downarrow} c^\dagger_{\alpha'\downarrow} c_{\alpha'\uparrow}\right)
\, ,
\end{eqnarray}
with $n_{i\alpha\sigma}$ being the number operator for the $t_{2g}$ orbitals
$\alpha \! \in \! \{a_{1g},e_g^{\pi+},e_g^{\pi-}\}$ \cite{KhomskiiZhETF}
at site $i,j \! \in \! \{1,2\}$ and \mbox{$\sigma \! \in \! \{\uparrow,\downarrow\}$}, 
and the operators $c^\dagger_{\alpha \sigma}$ create holes. 
We stick to $P6_3/mmc$ symmetry for the analysis of our RIXS data and consider 
the trigonal crystal field via $\Delta_{\rm CF} L_z^2$, splitting the $t_{2g}$ level into
$a_{1g}$ and $e_g^\pi$ orbitals, and the intersite hopping amplitudes $t_{a_{1g}}$ and 
$t_{e_g^\pi}$. For $D_{3h}$ symmetry of the Ir$_2$O$_9$ dimer, hopping is diagonal 
in the $a_{1g}$--$e_g^\pi$ basis.
Finally, we employ spin-orbit coupling $\zeta \, \mathbf{l}\cdot \mathbf{s}$.

\begin{figure}[t]
	\centering
	\includegraphics[width=\columnwidth]{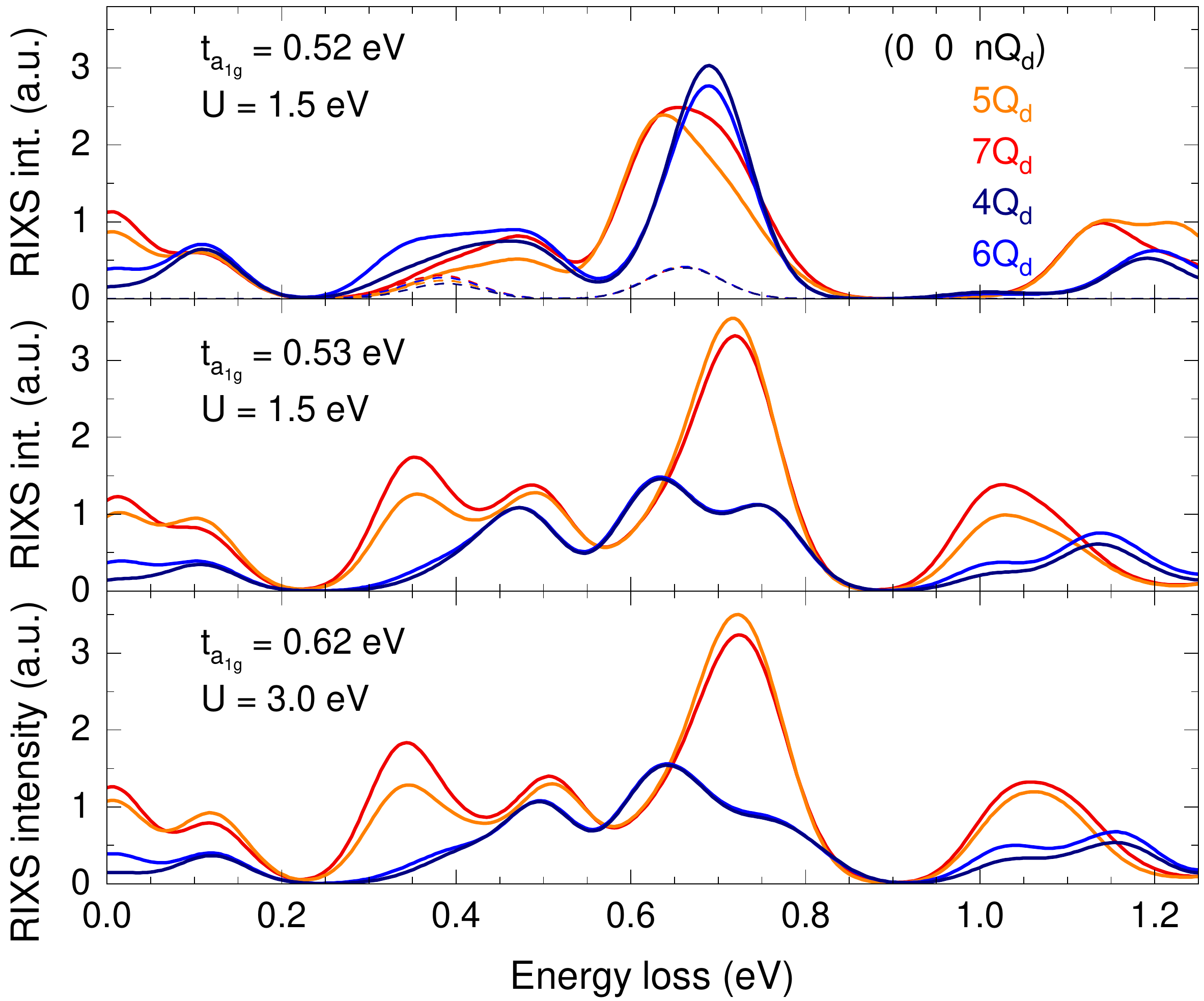}
	\caption{Exact diagonalization results for RIXS spectra at (0\,\,0\,\,$nQ_d$).  
		Top and middle panels show data on opposite sides of the phase transition for 
		$t_{a_{1g}}$\,=\,0.52\,eV and 0.53\,eV, respectively. 
		All other parameters are identical, i.e., $\zeta$\,=\,0.43\,eV, 
		$U$\,=\,1.5\,eV, $J_H$\,=\,0.33\,eV, $\Delta_{\rm CF}$\,=\,0.1\,eV,
		and $t_{e_g^\pi}/t_{a_{1g}}$\,=\,$-0.6$, as used in the middle panel of Fig.\ \ref{fig:3holeQMOs}.
		Bottom: result for larger $t_{a_{1g}}$\,=\,0.62\,eV and $U$\,=\,3.0\,eV.\@ 
		To take into account the finite In-Ir site disorder, all spectra contain the small contribution 
		of individual $t_{2g}^4$ sites depicted by dashed lines in the top panel (see main text), using the parameters given above. 
	}
	\label{fig:theoRIXS}
\end{figure}

Discussing three $t_{2g}$ holes on two Ir sites,
we start from vanishing hopping $t$\,=\,$t_{a_{1g}}$\,=\,$t_{e_g^\pi}$\,=\,0 and a 
cubic crystal field, i.e., we neglect the non-cubic part by choosing $\Delta_{\rm CF}$\,=\,$0$.
States with all three holes on the same site are about 2\,$U$ above the ground state and can be neglected.
For $t$\,=\,0 we hence are facing simple product states $|t_{2g}^5\rangle_i\,|t_{2g}^4\rangle_{j}$
of single-site multiplets with one and two holes. 
Their joint excitation energies as function of $\zeta$ are depicted in the left panel of 
Fig.\ \ref{fig:3holeQMOs} for $U$\,=\,1.5\,eV and $J_H$\,=\,0.33\,eV.\@ 
Energies are plotted up to $\zeta$\,=\,0.43\,eV, a realistic value for 
iridates \cite{KimKhalNa14}.  
While $t_{2g}^5$ systems in cubic approximation show a single intra-$t_{2g}$ excitation 
at $3/2\zeta \approx 0.7$\,eV \cite{Gretarsson13,Rossi17,Liu12,Revelli19b,Reig20,Khan21,Moretti14}
the dominant RIXS features in $t_{2g}^4$ compounds are observed 
at roughly 0.4 and 0.7\,eV \cite{Yuan17,Kusch18,Nag18}, see dashed lines in top panel of 
Fig.\ \ref{fig:theoRIXS}.

Hopping is diagonal in the single-particle $a_{1g}$--$e_g^\pi$ basis and mixes the local multiplets. 
We therefore skip a more detailed discussion of the $t_{2g}^4$ multiplets but mention in passing 
that the local $t_{2g}^4$ ground state shows $j$\,=\,0.
In contrast to the $t_{2g}^4$ states, the single-hole $t_{2g}^5$ states with $j$\,=\,1/2 and 3/2 
are still important for an intuitive picture of the electronic states of a dimer. 
In particular, most of the characteristic features of the electronic structure already 
become apparent when considering an idealized system with $\Delta_{\rm CF}$\,=\,$J_H$\,=\,0 and 
$t_{e_g^\pi}/t_{a_{1g}}$\,=\,$-1$. The latter makes hopping block-diagonal in the $j$\,=\,1/2 
and 3/2 basis, so that $j$ (but not $j_z$) remains a good quantum number.  
In this case, the eigenstates can be built from bonding and antibonding states 
\begin{align}\label{eq:BAB}
|j^m_\pm\rangle &\equiv \frac{1}{\sqrt{2}} \big( |j,m\rangle_1  \pm |j,m\rangle_2 \big) \, ,
\end{align}
where $m$ denotes suitable linear combinations of 
$j_z$\,=\,$\pm 1/2$ or $j_z$\,=\,$\pm 3/2, \pm 1/2$, 
and the definition of states on sites 1 and 2 takes into account that 
the IrO$_6$ are rotated with respect to each other \cite{KhomskiiZhETF}.
Product states of these simple states form a very good starting point for the 
discussion of the many-body eigenstates, even though $t_{e_g^\pi}/t_{a_{1g}}$ 
is closer to $-1/2$ in Ba$_3$InIr$_2$O$_9$ and Coulomb interaction is strong.

For the idealized case with $t$\,=\,$t_{a_{1g}}$\,=\,$-t_{e_g^\pi}$, 
the right panel of Fig.\ \ref{fig:3holeQMOs} depicts the effect of hopping 
on the excitation energies. We employ $\zeta$\,=\,0.43\,eV and $U$\,=\,1.5\,eV, 
keeping $\Delta_{\rm CF}$\,=\,$J_H$\,=\,0 for simplicity. 
For not too large $t$, the doubly degenerate ground state shows two holes in the bonding 
$j$\,=\,1/2 orbital $|\sfrac{1}{2}^\sigma_+\rangle$ and the third hole in the antibonding orbital, 
resulting in $j_{\rm dim}$\,=\,1/2, 
\begin{align}\label{eq:0}
|\sfrac{1}{2}_{\rm dim}\rangle_\sigma 
& =  |\sfrac{1}{2}^\uparrow_+\rangle \, |\sfrac{1}{2}^\downarrow_+\rangle \,
 |\sfrac{1}{2}^\sigma_-\rangle\, , 
\end{align}
as sketched in the left panel of Fig.\ \ref{fig:sketch}. 
In contrast, the first excited state at small $t$ has two holes in the antibonding orbital, 
\begin{align}\label{eq:1}
|1\rangle_\sigma & =  |\sfrac{1}{2}^\sigma_+\rangle \, |\sfrac{1}{2}^\uparrow_-\rangle \,
|\sfrac{1}{2}^\downarrow_-\rangle\, .
\end{align}
The absolute energies of 
$|\sfrac{1}{2}_{\rm dim}\rangle_\sigma$ 
and $|1\rangle_\sigma$ vary like $\pm t$, and hence the 
excitation energy of $|1\rangle_\sigma$ rises with a slope $2t$. Even though $U$ is large, 
this perfectly agrees with a simple picture of non-interacting holes since hopping for 
these three-particle states does not affect double occupancy.

The further low-energy states show one hole in an (anti-) bonding $j$\,=\,3/2 orbital, 
\begin{align}\label{eq:2}
|\sfrac{3}{2}_{\rm dim}\rangle_m 
& =  |\sfrac{1}{2}^\uparrow_+\rangle \, |\sfrac{1}{2}^\downarrow_+\rangle \,
|\sfrac{3}{2}^m_+\rangle \, 
\\ \label{eq:3}
|2\rangle_m 
& =  |\sfrac{1}{2}^\uparrow_+\rangle \, |\sfrac{1}{2}^\downarrow_+\rangle \,
|\sfrac{3}{2}^m_-\rangle \, 
\\ \label{eq:4}
|3\rangle_{\sigma\sigma^\prime m} 
& =  |\sfrac{1}{2}^\sigma_+\rangle \, |\sfrac{1}{2}^{\sigma^\prime}_-\rangle \,
|\sfrac{3}{2}^m_+\rangle \, . 
\end{align}
For $t$\,=\,0, all of these states lie at $\sfrac{3}{2}\, \zeta$. For $U$\,=\,0, their energies 
are given in the simple non-interacting picture described above, i.e., each bonding (antibonding) 
state contributes $-t$ ($+t$) to the excitation energy. In this limit, the energy of state 
$|\sfrac{3}{2}_{\rm dim}\rangle_m$ decreases with a slope $-3t$, which corresponds to an 
excitation energy $\sfrac{3}{2}\, \zeta -2t$. Accordingly, $|\sfrac{3}{2}_{\rm dim}\rangle_m$ 
becomes the ground state for $t$ larger than a critical value, 
as sketched in the right panel of Fig.\ \ref{fig:sketch}. 
For finite $U$, the slope is reduced and the behavior is not perfectly linear anymore, 
see right panel of Fig.\ \ref{fig:3holeQMOs}. 
The reduced slope reflects the Coulomb interaction between holes in the bonding orbitals 
for $j$\,=\,1/2 and 3/2 which compete for the same space between the two Ir sites.

In contrast to the behavior of $|\sfrac{3}{2}_{\rm dim}\rangle_m$, the energies of 
$|2\rangle_m$ and $|3\rangle_{\sigma\sigma^\prime m}$ vary like $-t$, independent of $U$. 
Therefore, their excitation energy remains constant, $\sfrac{3}{2}\, \zeta$. 
With this we collected the main ingredients to describe the general properties observed 
in the RIXS spectra of Ba$_3$InIr$_2$O$_9$. For realistic parameters, the simple states 
described above are split into a multitude of levels, see middle panel of Fig.\ \ref{fig:3holeQMOs}, 
but the main features of the RIXS spectra can still be explained in an intuitive picture, 
as described in the following and depicted in Fig.\ \ref{fig:sketch}.

First of all, the insensitivity of the excitation energy to hopping roughly remains valid 
for many states. This applies to the phase observed for small hopping but still is valid 
above the phase transition. This explains that the main RIXS intensity in Ba$_3$InIr$_2$O$_9$ 
is found in the energy range where single-site-like $d^4$ and $d^5$ iridates show RIXS features 
based on $\zeta$ and $J_H$ 
\cite{Gretarsson13,Rossi17,Liu12,Revelli19b,Reig20,Khan21,Moretti14,Yuan17,Kusch18,Nag18}, 
i.e., around 0.4\,eV and 0.7\,eV.\@ 
Note that this does not imply that the character of these states is the same as for vanishing hopping. 
It rather signifies that these states 
in terms of bonding/antibonding have the same character as the ground state. 
For instance for small hopping, they are mainly built  from two bonding and one antibonding orbital 
and hence show a very similar dependence on hopping as the state $|\sfrac{1}{2}_{\rm dim}\rangle_\sigma$.

Second, the observation of distinct low-energy RIXS features below 0.3\,eV suggests that 
Ba$_3$InIr$_2$O$_9$ is close to the phase transition \cite{Li20} where 
-- in the simple picture given in the right panel of Fig.\ \ref{fig:3holeQMOs} 
and sketched in Fig.\  \ref{fig:sketch}  -- the ground state changes from 
$|\sfrac{1}{2}_{\rm dim}\rangle_\sigma$ to $|\sfrac{3}{2}_{\rm dim}\rangle_m$. 
The energies of the low-energy features strongly disperse as a function of hopping and 
hence restrict the possible range of $t_{a_{1g}}$  and $t_{e_g^\pi}$. However, the slope 
of these states is renormalized by Coulomb interaction, as discussed above. 
Therefore the relevant range of hopping parameters depends on $U$ \cite{Li20}. 
Deviations from the idealized case $t_{e_g^\pi}/t_{a_{1g}}$\,=\,$-1$ 
mix states with different $j$ and thereby lift the four-fold degeneracy in 
Eq.\ \eqref{eq:2}, giving rise to two Kramers doublets. 
For the parameters used in the middle panel of Fig.\ \ref{fig:3holeQMOs}, 
this splitting is of the same order of magnitude as the non-cubic crystal-field 
splitting  $\Delta_{\rm CF}$.

With Ba$_3$InIr$_2$O$_9$ being close to the phase transition, the central question to be answered 
concerns the ground state. We address this issue via the RIXS matrix elements and calculate 
RIXS spectra in the dipole approximation as described in Ref.\ \cite{Revelli19}. 
In the top and middle panels of Fig.\  \ref{fig:theoRIXS}, we compare results for 
$t_{a_{1g}}$\,=\,0.52\,eV and 0.53\,eV, located on opposite sides of the phase transition. 
All other parameters are kept fixed. In contrast to the excitation energies that are very similar 
in the two cases, the RIXS matrix elements sensitively depend on the ground state, which is 
reflected in the RIXS intensity of different features as well as in their dependence on $q_c$.
In the experiment, we find that the RIXS intensity is larger for odd values of $Q_d$ (red and 
orange curves in Figs.\ \ref{fig:rixst2geg} and \ref{fig:theoRIXS}). This is reproduced for 
hopping larger than the critical value, i.e., a ground state related to 
$|\sfrac{3}{2}_{\rm dim}\rangle_m$. 
The corresponding middle panel of Fig.\ \ref{fig:theoRIXS} shows reasonable agreement with 
the experimental data concerning the main features of the spectrum. In contrast, the overall 
RIXS intensity is very similar for even and odd $Q_d$ for $t_{a_{1g}}$\,=\,0.52\,eV, 
i.e., smaller than the critical value where the ground state is based on 
$|\sfrac{1}{2}_{\rm dim}\rangle_\sigma$. 
We hence conclude that hopping exceeds the critical value while the presence of low-energy 
features demonstrates that Ba$_3$InIr$_2$O$_9$ is still close to the phase transition, 
see Fig.\ \ref{fig:sketch}.
Concerning the character of the ground state, our result agrees with 
an exact diagonalization study by Li \textit{et al.} \cite{Li20} which finds 
evidence for a $j_{\rm dim}$\,=\,3/2 nature of the ground state of Ba$_3$InIr$_2$O$_9$ 
based on the small anisotropy of the magnetic susceptibility and the small magnitude of 
the effective magnetic moments. Furthermore, the corresponding sequence 
with the lowest bonding $j$\,=\,3/2 orbital lying below the antibonding $j$\,=\,1/2 orbital 
agrees with a RIXS study on Ba$_3$CeIr$_2$O$_9$ with two holes per dimer \cite{Revelli19}.

Concerning the electronic parameters, we use $\zeta$\,=\,0.43\,eV and $J_H$\,=\,0.33\,eV, 
which are typical for iridates \cite{KimKhalNa14,Revelli19}.
Determination of the hopping parameters is more subtle. For given $\zeta$, the critical hopping for 
the phase transition depends on $U$ \cite{Li20}. Using $t_{e_g^\pi}/t_{a_{1g}}$\,=\,$-0.6$, 
we obtain very similar results for the RIXS spectra for $U$\,=\,1.5\,eV and 3.0\,eV 
if $t_{a_{1g}}$ is adapted from 0.53\,eV to 0.62\,eV, see Fig.\ \ref{fig:theoRIXS}. 
These values roughly agree with first-principles density-functional theory calculations \cite{YingLi}
which yield $t_{a_{1g}}$\,=\,$0.47$\,eV and $t_{e_g^\pi}/t_{a_{1g}}$=\,$-0.53$.
Finally, the effect of a small crystal-field splitting $\Delta_{\rm CF}$ is not pronounced, 
which prevents an accurate determination. We employ $\Delta_{\rm CF}$\,=\,0.1\,eV.\@

Both finite $\Delta_{\rm CF}$ and $t_{e_g^\pi}/t_{a_{1g}}$\,$\neq$\,$-1$ split the 
quartet $|\sfrac{3}{2}_{\rm dim}\rangle_m$ into two doublets. 
For the parameters derived above, the lower-energy doublet forms the many-body ground state, 
while the first excited state is related to $|\sfrac{1}{2}_{\rm dim}\rangle_\sigma$,   
see middle panel of Fig.\ \ref{fig:3holeQMOs}. 
We emphasize that 86\,\% of the weight of the ground-state doublet arises from the 
quartet $|\sfrac{3}{2}_{\rm dim}\rangle_m$. 
The main admixture, carrying about 8\,\%, stems from 
\begin{align}\label{eq:2prime}
|4\rangle_m
& =  |\sfrac{1}{2}^\uparrow_-\rangle \, |\sfrac{1}{2}^\downarrow_-\rangle \,
|\sfrac{3}{2}^{m}_+\rangle \, 
\\
|5\rangle_{\sigma\sigma^\prime m} 
& =  |\sfrac{1}{2}^\sigma_+\rangle \, |\sfrac{1}{2}^{\sigma^\prime}_-\rangle \,
|\sfrac{3}{2}^{m}_-\rangle \, , 
\end{align}
and a further 3\,\% are contributed by 
\begin{align}\label{eq:6}
|6\rangle_{\sigma m m^\prime}
	& =  |\sfrac{1}{2}^\sigma_+\rangle \, |\sfrac{3}{2}^m_+\rangle \,
	|\sfrac{3}{2}^{m^\prime}_+\rangle \, . 
\end{align}
Similarly, $|\sfrac{1}{2}_{\rm dim}\rangle_\sigma$ carries about 83\,\% of the weight 
of the lowest excited many-body state, and $|2\rangle_m$ and 
$|3\rangle_{\sigma\sigma^\prime m}$ together contribute 8\,\%. 
We hence conclude that the spin-orbit entangled bonding and antibonding orbitals capture 
the dominant physics for realistic parameters.

Finally, we address the possible role of In-Ir site disorder on the RIXS spectra. 
With one exception, the exact diagonalization calculations describe 
the experimentally observed peak energies very well, see red symbols in the middle panel 
of Fig.\ \ref{fig:3holeQMOs}. At low energy, however, theory predicts two excitation energies, 
and our calculations do not yield any RIXS peak around 0.2\,eV.\@ 
This can be reconciled with the experimental data assuming that the weak RIXS intensity 
around 0.23\,eV, cf.\ Fig.\ \ref{fig:rixs3} and blue symbol in Fig.\ \ref{fig:3holeQMOs}, 
is caused by a magnetic contribution of a few \% of Ir ions on In sites
that form roughly 180$^\circ$ Ir-O-Ir bonds with Ir on dimer sites, 
as discussed in Sec.\ \ref{sec:lowE}.
Furthermore, the calculated spectra contain a small contribution of individual 
$t_{2g}^4$ Ir$^{5+}$ sites, see dashed lines in top panel of Fig.\ \ref{fig:theoRIXS}. 
These reflect the presence of 7.3\,\% In$^{3+}$ ions on dimer sites, as derived 
from our analysis of the crystal structure, see Sec.\ \ref{sec:struc}. 
For simplicity, we assume that an In$^{3+}$ ion on a dimer site is accompanied by 
an Ir$^{5+}$ ion. This contribution, however, is small and overlaps with dimer features, 
it is hence not important for our analysis. 
Around 0.4\,eV, the small $\mathbf{q}$ dependence of the intensity of this single-site 
contribution reflects a corresponding change of the scattering geometry. Such effects 
are typically much smaller than the $\mathbf{q}$ dependence observed for the 
quasimolecular dimer contribution.

\section{Conclusions}

In spin-orbit entangled iridates, experimental realizations of quantum spin liquids are very limited
and hotly debated. The honeycomb iridate H$_3$LiIr$_2$O$_6$ for instance does not show
magnetic order down to 50\,mK \cite{Kitagawa18} but the role of disorder related to the H ions
has been pointed out \cite{Yadav18,Li18,Knolle19b}. 
The compound Ba$_3$InIr$_2$O$_9$ was discussed as an outstanding case with clear experimental evidence
for persistent spin dynamics down to 20\,mK from thermodynamic and spectroscopic data \cite{Dey17}. 
The dimer structure based on face-sharing IrO$_6$ octahedra requires solving the issue of the
microscopic character of the local moments on a quantitative level. Understanding these local moments
is a prerequisite for a successful description of magnetism in this compound.

Overall, our results establish that Ba$_3$InIr$_2$O$_9$ is a spin-orbit-entangled cluster Mott insulator. 
We prove the insulating character via dielectric spectroscopy and demonstrate 
that the crystal structure shows a monoclinic distortion already at room temperature. 
Based on the observation of a double-slit type sinusoidal interference pattern, our RIXS data
establish the quasimolecular orbital character of the electronic structure in
face-sharing geometry. The three holes are fully delocalized over the two sites of a dimer. 
The physics is related to the case of the sister compound Ba$_3$CeIr$_2$O$_9$ in which two holes 
per dimer form a spin-orbit entangled nonmagnetic $j_{\rm dim}$\,=\,0 ground state \cite{Revelli19}.
In Ba$_3$InIr$_2$O$_9$, the many-body dimer ground state can be approximated as a spin-orbit entangled 
Kramers doublet based on the $j_{\rm dim}$\,=\,3/2 state in which two holes occupy the 
bonding orbital built from local $j\,$=\,1/2 moments, as in Ba$_3$CeIr$_2$O$_9$, while the third hole 
is in a bonding $j$\,=\,3/2 state. This character of the ground state explains the dominant 
$\sin^2(q_c\,d/2)$ behavior of the RIXS intensity. 
The simple picture of bonding orbitals of local $j$ states works best in the idealized case 
$t_{e_g^\pi}/t_{a_{1g}}$\,=\,$-1$, where hopping is block-diagonal in the $j$ basis, 
but it still yields a reasonable description for realistic parameters. 
Furthermore, the observation of low-energy RIXS features demonstrates that Ba$_3$InIr$_2$O$_9$ 
is close to the phase transition to a $j_{\rm dim}$\,=\,1/2 state \cite{Li20}.

The realization of a new type of a gapless spin liquid based on spin-orbit entangled 
moments in quasimolecular orbitals would open up a different and very promising perspective 
on iridates and quantum spin liquids in general. In Ba$_3$InIr$_2$O$_9$, the proposed 
unconventional spin-liquid character is supported by a quadratic behavior found in the specific heat 
at low temperature, which deviates from the typical spin-liquid scenario of fermionic 
spinons \cite{Dey17}. However, this has to be taken with a grain of salt since the 
possible role of a small percentage of cation disorder needs to be further elucidated.

The magnetism in Ba$_3$InIr$_2$O$_9$ is based on interdimer exchange interactions which 
occur on an energy scale that challenges the state-of-the-art energy resolution of RIXS at 
the Ir $L_3$ edge, calling for thorough theoretical studies of exchange interactions between 
quasimolecular states on neighboring dimers and further spectroscopic investigations 
with other techniques.

Finally, our results demonstrate the promising potential of 
RIXS interference patterns $I(\mathbf{q})$ to unravel the electronic structure 
and to determine the character and symmetry of electronic states \cite{Gelmukhanov94,Ma94,Ma95}. 
Using the iridate dimers Ba$_3$CeIr$_2$O$_9$ \cite{Revelli19} and Ba$_3$InIr$_2$O$_9$ 
as well-defined model systems, the power of this approach has been established. 
Furthermore, such interference patterns revealed the nearest-neighbor character of 
magnetic excitations in the honeycomb iridates with dominant Kitaev exchange \cite{Revelli20}. 
However, the technique is expected to apply equally well to trimers or larger clusters 
as well as to ladders, bilayers, and other superstructures. 
Analyzing only a few values of $\mathbf{q}$, this interference effect has also been addressed 
in the bilayer iridate Sr$_3$Ir$_2$O$_7$ \cite{MorettiSr3} as well as in VO$_2$ \cite{He16} 
and in the context of stripes in nickelates and cuprates \cite{Schuelke11}. 
One advantage of the iridates, compared to $4d$ or $3d$ transition metal compounds, 
is that the Ir $L$ edge is lying in the hard x-ray region, allowing us to cover a large range 
of $\mathbf{q}$ space and hence to observe the interference pattern over more than a period.

\acknowledgments

We thank Y. Li, S. M. Winter, and R. Valent{\'\i} for fruitful discussions
and for sharing their DFT results with us.
We acknowledge funding from the Deutsche Forschungsgemeinschaft (DFG, German Research Foundation)
-- Project numbers 277146847 and 247310070 -- CRC 1238 (projects A02, B02, B03, and C03)
and CRC 1143 (project A05), respectively. 
The work in Augsburg was supported by the DFG through project 107745057 (TRR 80) and
by the Federal Ministry for Education and Research
through the Sofja Kovalevskaya Award of the Alexander von Humboldt Foundation.
M.H.\@ acknowledges partial funding by the Knut and Alice Wallenberg Foundation and the Swedish Research Council.


\begin{thebibliography}{99}

\bibitem{WitczakKrempa14}
W. Witczak-Krempa, G. Chen, Y. B. Kim, and L. Balents,
\textit{Correlated Quantum Phenomena in the Strong Spin-Orbit Regime},
Annu. Rev. Condens. Matter Phys. \textbf{5}, 57 (2014).

\bibitem{Rau16}
J. G. Rau, E. K.-H. Lee, and H.-Y. Kee,
\textit{Spin-Orbit Physics Giving Rise to Novel Phases in Correlated Systems: Iridates and Related Materials},
Annu. Rev. Condens. Matter Phys. \textbf{7}, 195 (2016).

\bibitem{Schaffer16}
R. Schaffer, E. K.-H. Lee, B.-J. Yang, and Y. B. Kim,
\textit{Recent progress on correlated electron systems with strong spin-orbit coupling},
Rep. Prog. Phys. \textbf{79}, 094504 (2016).

\bibitem{Trebst17}
S. Trebst,
\textit{Kitaev Materials}, arXiv:1701.07056 (2017).

\bibitem{Winter17}
S.M. Winter, A.A. Tsirlin, M. Daghofer, J. van den Brink, Y. Singh, P. Gegenwart, and R. Valent{\'\i},
\textit{Models and materials for generalized Kitaev magnetism},
J. Phys.: Condens. Matter \textbf{29}, 493002 (2017).

\bibitem{Hermanns18}
M. Hermanns, I. Kimchi, and J. Knolle,
\textit{Physics of the Kitaev Model: Fractionalization, Dynamic Correlations, and Material Connections},
Annu. Rev. Condens. Matter Phys. \textbf{9}, 17 (2018).

\bibitem{Cao18}
G. Cao and P. Schlottmann,
\textit{The Challenge of Spin-Orbit-Tuned Ground States in Iridates},
Rep. Prog. Phys. \textbf{81}, 042502 (2018).

\bibitem{Takagi19}
H. Takagi, T. Takayama, G. Jackeli, G. Khaliullin and S. E. Nagler,
\textit{Concept and realization of Kitaev quantum spin liquids},
Nature Reviews Physics \textbf{1}, 264 (2019).

\bibitem{Motome20}
Y. Motome and J. Nasu,
\textit{Hunting Majorana Fermions in Kitaev Magnets},
J. Phys. Soc. Jpn. \textbf{89}, 012002 (2020).

\bibitem{Takayama21}
T. Takayama, J. Chaloupka, A. Smerald, G. Khaliullin, and H. Takagi,
\textit{Spin–Orbit-Entangled Electronic Phases in $4d$ and $5d$ Transition-Metal Compounds},
J. Phys. Soc. Jpn. \textbf{90}, 062001 (2021).

\bibitem{Nguyen21}
L. T. Nguyen and R. J. Cava,
\textit{Hexagonal Perovskites as Quantum Materials},
Chem. Rev. \textbf{121}, 2935 (2021).

\bibitem{Trebst22}
S. Trebst and C. Hickey,
\textit{Kitaev materials},
Phys. Rep. \textbf{950}, 1 (2022).

\bibitem{Jackeli09}
G. Jackeli and G. Khaliullin,
\textit{Mott Insulators in the Strong Spin-Orbit Coupling Limit: From Heisenberg to a Quantum Compass
	and Kitaev Models},
Phys. Rev. Lett. \textbf{102}, 017205 (2009).

\bibitem{Singh10}
Y. Singh and P. Gegenwart,
\textit{Antiferromagnetic Mott insulating state in single crystals of the honeycomb lattice material Na$_2$IrO$_3$},
Phys. Rev. B \textbf{82}, 064412 (2010).

\bibitem{Singh12}
Y. Singh, S. Manni, J. Reuther, T. Berlijn, R. Thomale, W. Ku, S. Trebst, and P. Gegenwart,
\textit{Relevance of the Heisenberg-Kitaev Model for the Honeycomb Lattice Iridates $A_2$IrO$_3$},
Phys. Rev. Lett. \textbf{108}, 127203 (2012).

\bibitem{Kitagawa18}
K. Kitagawa, T. Takayama, Y. Matsumoto, A. Kato, R. Takano, Y. Kishimoto, S. Bette, R. Dinnebier,
G. Jackeli, and H. Takagi,
\textit{A spin-orbital-entangled quantum liquid on a honeycomb lattice},
Nature \textbf{554}, 341 (2018).

\bibitem{KhomskiiZhETF}
D.I. Khomskii, K.I. Kugel, A.O. Sboychakov, and S.V. Streltsov,
\textit{Role of Local Geometry in the Spin and Orbital Structure of Transition Metal Compounds},
J. Exp. Theo. Phys. \textbf{122}, 484 (2016).

\bibitem{Doi04}
Y. Doi and Y. Hinatsu,
\textit{The structural and magnetic characterization of $6H$-perovskite-type oxides Ba$_3$$Ln$Ir$_2$O$_9$
	($Ln$ = Y, lanthanides)},
J. Phys.: Condens. Matter \textbf{16}, 2849  (2004).

\bibitem{Sakamoto06}
T. Sakamoto, Y. Doi, and Y. Hinatsu,
\textit{Crystal structures and magnetic properties of $6H$-perovskite-type oxides
	Ba$_3$$M$Ir$_2$O$_9$ ($M$\,=\,Mg, Ca, Sc, Ti, Zn, Sr, Zr, Cd and In)},
J. Sol. State Chem. \textbf{179}, 2595 (2006).

\bibitem{Dey12}
T. Dey, A.V. Mahajan, P. Khuntia, M. Baenitz, B. Koteswararao, and F.C. Chou,
\textit{Spin-liquid behavior in $J_{\rm eff}$\,=\,$1/2$ triangular lattice compound Ba$_3$IrTi$_2$O$_9$},
Phys. Rev. B \textbf{86}, 140405(R) (2012).

\bibitem{Kumar16}
R. Kumar, D. Sheptyakov, P. Khuntia, K. Rolfs, P.G. Freeman, H.M. R{\o}nnow,
T. Dey, M. Baenitz, and A.V. Mahajan,
\textit{Ba$_3$$M_x$Ti$_{3-x}$O$_9$ ($M$ = Ir, Rh): A family of $5d/4d$-based diluted quantum spin liquids},
Phys. Rev. B \textbf{94}, 174410 (2016).

\bibitem{Lee17}
W.-J. Lee, S.-H. Do, S. Yoon, S. Lee, Y.S. Choi, D.J. Jang, M. Brando, M. Lee, E.S. Choi, S. Ji,
Z.H. Jang, B.J. Suh, and K.-Y. Choi,
\textit{Putative spin liquid in the triangle-based iridate Ba$_3$IrTi$_2$O$_9$}, 
Phys. Rev. B \textbf{96}, 014432 (2017).

\bibitem{Kim04}
S.J. Kim, M.D. Smith, J. Darriet, and H.-C. zur Loye,
\textit{Crystal growth of new perovskite and perovskite related iridates:
	Ba$_3$LiIr$_2$O$_9$, Ba$_3$NaIr$_2$O$_9$, and Ba$_{3.44}$K$_{1.56}$Ir$_2$O$_{10}$},
J. Solid State Chem. \textbf{177}, 1493 (2004).

\bibitem{Dey14}
T. Dey, R. Kumar, A. V. Mahajan, S. D. Kaushik, and V. Siruguri,
\textit{Unconventional magnetism in the spin-orbit-driven Mott insulators Ba$_3M$Ir$_2$O$_9$ ($M$=Sc,Y)},
Phys. Rev. B \textbf{89}, 205101 (2014).

\bibitem{Dey17}
T. Dey, M. Majumder, J.C. Orain, A. Senyshyn, M. Prinz-Zwick, S. Bachus, Y. Tokiwa, F. Bert,
P. Khuntia, N. B\"{u}ttgen, A.A. Tsirlin, and P. Gegenwart,
\textit{Persistent low-temperature spin dynamics in the mixed-valence iridate Ba$_3$InIr$_2$O$_9$},
Phys. Rev. B \textbf{96}, 174411 (2017).

\bibitem{Khan19}
M. S. Khan, A. Bandyopadhyay, A. Nag, V. Kumar, A. V. Mahajan, and S. Ray,
\textit{Magnetic ground state of the distorted $6H$ perovskite Ba$_3$CdIr$_2$O$_9$},
Phys. Rev. B \textbf {100}, 064423 (2019).

\bibitem{Nag19}
A. Nag, S. Bhowal, M. Moretti Sala, A. Efimenko, I. Dasgupta, and S. Ray, 
\textit{Hopping-Induced Ground-State Magnetism in $6H$ Perovskite Iridates}, 
Phys. Rev. Lett. \textbf{123}, 017201 (2019).

\bibitem{Kumar21}
S. Kumar, S. K. Panda, M. M. Patidar, S. K. Ojha, P. Mandal, G. Das, J. W. Freeland,
V. Ganesan, P. J. Baker, and S. Middey,
\textit{Spin-liquid behavior of the three-dimensional magnetic system Ba$_3$NiIr$_2$O$_9$ with S=1},
Phys. Rev. B \textbf{103}, 184405 (2021).

\bibitem{Garg21}
C. Garg, D. Roy, M. Lonsky, P. Manuel, A. Cervellino, J. Müller, M. Kabir, and S. Nair,
\textit{Evolution of the structural, magnetic, and electronic properties of the triple
	perovskite Ba$_3$CoIr$_2$O$_9$},
Phys. Rev. B \textbf{103}, 014437 (2021).

\bibitem{Xu19}
L. Xu, R. Yadav, V. Yushankhai, L. Siurakshina, J. van den Brink, and L. Hozoi,
\textit{Superexchange interactions between spin-orbit-coupled $j \approx 1/2$ ions
	in oxides with face-sharing ligand octahedra},
Phys. Rev. B \textbf{99}, 115119 (2019).

\bibitem{Streltsov16}
S.V. Streltsov and D.I. Khomskii,
\textit{Covalent bonds against magnetism in transition metal compounds},
Proc. Nat Acad. Science \textbf{113}, 10491 (2016).

\bibitem{Mazin12}
I.I. Mazin, H.O. Jeschke, K. Foyevtsova, R. Valent{\'\i}, and D.I. Khomskii,
\textit{Na$_2$IrO$_3$ as a Molecular Orbital Crystal},
Phys. Rev. Lett. \textbf{109}, 197201 (2012).

\bibitem{Foy13}
K. Foyevtsova, H.O. Jeschke, I.I. Mazin, D.I. Khomskii, and R. Valent{\'\i},
\textit{Ab initio} analysis of the tight-binding parameters and magnetic interactions in Na$_2$IrO$_3$,
Phys. Rev. B \textbf{88}, 035107 (2013).

\bibitem{Wang18}
Y. Wang, Ruitang Wang, Jungho Kim, M. H. Upton, D. Casa, T. Gog, G. Cao, G. Kotliar, M. P. M. Dean, and X. Liu,
\textit{Direct detection of dimer orbitals in Ba$_5$AlIr$_2$O$_{11}$},
Phys. Rev. Lett. \textbf{122}, 106401 (2019).

\bibitem{Streltsov17}
S.V. Streltsov, G. Cao, and D.I. Khomskii,
\textit{Suppression of magnetism in Ba$_5$AlIr$_2$O$_{11}$: Interplay of Hund's coupling, molecular orbitals,
	and spin-orbit interaction},
Phys. Rev. B \textbf{96}, 014434 (2017).

\bibitem{Miiller12}
W. Miiller, M. Avdeev, Q. Zhou, B.J. Kennedy, N. Sharma, R. Kutteh, G.J. Kearley, S. Schmid,
K.S. Knight, P.E.R. Blanchard, and C.D. Ling,
\textit{Giant Magnetoelastic Effect at the Opening of a Spin-Gap in Ba$_3$BiIr$_2$O$_9$},
J. Amer. Chem. Soc. \textbf{134}, 3265 (2012).

\bibitem{Revelli19}
A. Revelli, M. Moretti Sala, G. Monaco, P. Becker, L. Bohat\'{y}, M. Hermanns, T.C.~Koethe,
T. Fr\"{o}hlich, P. Warzanowski, T. Lorenz, S.V. Streltsov, P.H.M. van Loosdrecht, D.I. Khomskii,
J. van den Brink, and M. Gr\"{u}ninger,
\textit{Resonant inelastic x-ray incarnation of Young's double-slit experiment},
Sci. Adv. \textbf{5}, eaav4020 (2019).

\bibitem{Li20}
Y. Li, A.A. Tsirlin, T. Dey, P. Gegenwart, R. Valent{\'\i}, and S. M. Winter,
\textit{Soft and anisotropic local moments in $4d$ and $5d$ mixed-valence $M_2$O$_9$ dimers},
Phys. Rev. B \textbf{102}, 235142 (2020).

\bibitem{Khaliullin13}
G. Khaliullin, 
\textit{Excitonic Magnetism in Van Vleck–type $d^4$ Mott Insulators}, 
Phys. Rev. Lett. \textbf{111}, 197201 (2013).

\bibitem{KhomskiiBook}
D.I. Khomskii, \textit{Transition metal compounds}, 
Cambridge University Press (2014).

\bibitem{Kaushal19}
N. Kaushal, A. Nocera, G. Alvarez, A. Moreo, and E. Dagotto, 
\textit{Block excitonic condensate at $n$=3.5 in a spin-orbit coupled $t_{2g}$ 
	multiorbital Hubbard model},
Phys. Rev. B \textbf{99}, 155115 (2019).

\bibitem{AbdElmeguid04}
M. M. Abd-Elmeguid, B. Ni, D. I. Khomskii, R. Pocha, D. Johrendt, X. Wang, and K. Syassen,
\textit{Transition from Mott Insulator to Superconductor in GaNb$_4$Se$_8$ and GaTa$_4$Se$_8$
	under High Pressure},
Phys. Rev. Lett. \textbf{93}, 126403 (2004).

\bibitem{Chen14}
G. Chen, H.-Y. Kee, and Y. B. Kim,
\textit{Fractionalized Charge Excitations in a Spin Liquid on Partially Filled Pyrochlore Lattices},
Phys. Rev. Lett. \textbf{113}, 197202 (2014).

\bibitem{Khomskii21}
D. I. Khomskii and S. V. Streltsov, 
\textit{Orbital Effects in Solids: Basics, Recent Progress, and Opportunities}, 
Chem. Rev.  \textbf{121}, 2992 (2021).

\bibitem{Moretti18}
M. Moretti Sala, K. Martel, C. Henriquet, A. Al Zein, L. Simonelli, Ch.J. Sahle, H. Gonzalez,
M.-C. Lagier, C. Ponchut, S. Huotari, R. Verbeni, M. Krisch, and G. Monaco,
\textit{A high-energy-resolution resonant inelastic X-ray scattering spectrometer at ID20
	of the European Synchrotron Radiation Facility},
J. Synchrotron Rad. \textbf{25}, 580 (2018).

\bibitem{Moretti13}
M. Moretti Sala, C. Henriquet, L. Simonelli, R. Verbeni, and G. Monaco,
\textit{High energy-resolution set-up for Ir $L_3$ edge RIXS experiments},
J. Electron Spectrosc. Relat. Phenom. \textbf{188}, 150 (2013).

\bibitem{karplus}
P. A. Karplus and K. Diederichs,
\textit{Linking Crystallographic Model and Data Quality},
Science \textbf{336}, 1030 (2012).

\bibitem{petricek}
V. Pet\v{r}\'{i}\v{c}ek, M. Du\v{s}ek, and L. Palatinus,
\textit{Crystallographic Computing System JANA2006: General features},
Z. Kristallogr. \textbf{229}, 345 (2014).

\bibitem{Becker1974a}
P. J. Becker and P. Coppens,
\textit{Extinction within the Limit of Validity of the Darwin Transfer Equations},
Acta Cryst. \textbf{A30}, 129 (1974).

\bibitem{Stokes88}
H. T. Stokes and D. M. Hatch, 
\textit{Isotropy subgroups of the 230 crystallographic space groups} (World Scientific, 1988).

\bibitem{Cwik2003}
M. Cwik, T. Lorenz, J. Baier, R. M\"uller, G. Andr\'e, F. Bour\'ee, F. Lichtenberg, A. Freimuth,
R. Schmitz, E. M\"uller-Hartmann, and M. Braden,
\textit{Crystal and magnetic structure of LaTiO$_3$: Evidence for nondegenerate t$_{2g}$ orbitals}, 
Phys. Rev. B \textbf{68}, 060401(R) (2003).

\bibitem{Gretarsson13}
H. Gretarsson, J.P. Clancy, X. Liu, J.P. Hill, E. Bozin, Y. Singh, S. Manni, P. Gegenwart,
J. Kim, A.H. Said, D. Casa, T. Gog, M.H. Upton, H.-S. Kim, J. Yu, V.M. Katukuri,
L. Hozoi, J. van den Brink, and Y.-J. Kim,
\textit{Crystal-Field Splitting and Correlation Effect on the Electronic Structure of $A_2$IrO$_3$},
Phys. Rev. Lett. \textbf{110}, 076402 (2013).

\bibitem{Rossi17}
M. Rossi, M. Retegan, C. Giacobbe, R. Fumagalli, A. Efimenko, T. Kulka, K. Wohlfeld, A. I. Gubanov,
and M. Moretti Sala,
\textit{Possibility to realize spin-orbit-induced correlated physics in iridium fluorides},
Phys. Rev. B \textbf{95}, 235161 (2017).

\bibitem{Liu12}
X. Liu, V. M. Katukuri, L. Hozoi, Wei-Guo Yin, M. P. M. Dean, M. H. Upton, Jungho Kim, D. Casa, A. Said,
T. Gog, T. F. Qi, G. Cao, A. M. Tsvelik, J. van den Brink, and J. P. Hill,
\textit{Testing the Validity of the Strong Spin-Orbit-Coupling Limit for Octahedrally Coordinated Iridate
	Compounds in a Model System Sr$_3$CuIrO$_6$},
Phys. Rev. Lett. \textbf{109}, 157401 (2012).

\bibitem{Revelli19b}
A. Revelli, C.C. Loo, D. Kiese, P. Becker, T. Fr\"ohlich, T. Lorenz, M. Moretti Sala, G.~Monaco, F.L. Buessen,
J. Attig, M. Hermanns, S.V.~Streltsov, D.I. Khomskii, J. van den Brink, M. Braden, P.H.M. van Loosdrecht,
S. Trebst, A. Paramekanti, and M. Gr\"uninger,
\textit{Spin-orbit entangled $j$\,=\,$1/2$ moments in Ba$_2$CeIrO$_6$: A  frustrated fcc quantum magnet},
Phys. Rev. B \textbf{100}, 085139 (2019). 

\bibitem{Reig20}
D. Reig-i-Plessis, T. A. Johnson, K. Lu, Q. Chen, J. P. C. Ruff, M. H. Upton, 
T. J. Williams, S. Calder, H. D. Zhou, J. P. Clancy, A. A. Aczel, and G. J. MacDougall, 
\textit{Structural, electronic, and magnetic properties of nearly ideal $J_{eff}$\,=\,1/2 iridium halides}, 
Phys. Rev. Materials \textbf{4}, 124407 (2020).

\bibitem{Khan21}
N. Khan, D. Prishchenko, M.H. Upton, V.G. Mazurenko, and A.A. Tsirlin, 
\textit{Towards cubic symmetry for Ir$^{4+}$: Structure and magnetism of the antifluorite K$_2$IrBr$_6$}, 
Phys. Rev. B \textbf{103}, 125158 (2021).

\bibitem{Moretti14}
M. Moretti Sala, S. Boseggia, D. F. McMorrow, and G. Monaco, 
\textit{Resonant X-Ray Scattering and the $j_{eff}$\,=\,1/2 Electronic Ground State in Iridate Perovskites}, 
Phys. Rev. Lett. \textbf{112}, 026403 (2014).

\bibitem{Yuan17}
Bo Yuan, J. P. Clancy, A. M. Cook, C. M. Thompson, J. Greedan, G. Cao, B. C. Jeon, T. W. Noh, M. H. Upton,
D. Casa, T. Gog, A. Paramekanti, and Young-June Kim,
\textit{Determination of Hund’s coupling in 5d oxides using resonant inelastic x-ray scattering},
Phys. Rev. B \textbf{95}, 235114 (2017).

\bibitem{Kusch18}
M. Kusch, V. M. Katukuri, N. A. Bogdanov, B. B\"{u}chner, T. Dey, D. V. Efremov,
J. E. Hamann-Borrero, B. H. Kim, M. Krisch, A. Maljuk, M. Moretti Sala, S. Wurmehl,
G. Aslan-Cansever, M. Sturza, L. Hozoi, J. van den Brink, and J. Geck,
\textit{Observation of heavy spin-orbit excitons propagating in a nonmagnetic background:
	The case of (Ba,Sr)$_2$YIrO$_6$},
Phys. Rev. B \textbf{97}, 064421 (2018).
	
\bibitem{Nag18}
A. Nag, S. Bhowal, A. Chakraborty, M. M. Sala, A. Efimenko, F. Bert, P. K. Biswas, A. D. Hillier,
M. Itoh, S. D. Kaushik, V. Siruguri, C. Meneghini, I. Dasgupta, and Sugata Ray,
\textit{Origin of magnetic moments and presence of spin-orbit singlets in Ba$_2$YIrO$_6$},
Phys. Rev. B \textbf{98}, 014431 (2018).

\bibitem{Ament11}
L.J.P. Ament, M. van Veenendaal, T.P. Devereaux, J.P. Hill, and J. van den Brink,
\textit{Resonant inelastic x-ray scattering studies of elementary excitations},
Rev. Mod. Phys. \textbf{83}, 705 (2011).

\bibitem{Gelmukhanov94}
F. Gel'mukhanov and H. Agren,
\textit{Resonant inelastic x-ray scattering with symmetry-selective excitation},
Phys. Rev. A \textbf{49}, 4378 (1994).

\bibitem{Ma94}
Y. Ma,
\textit{X-ray absorption, emission, and resonant inelastic scattering in solids},
Phys. Rev. B \textbf{49}, 5799 (1994).

\bibitem{Ma95}
Y. Ma and M. Blume,
\textit{Interference of fluorescence x rays and coherent excitation of core levels,}
Rev. Sci. Instr. \textbf{66}, 1543 (1995).

\bibitem{Revelli20}
A. Revelli, M. Moretti Sala, G. Monaco, C. Hickey, P. Becker, F. Freund, A. Jesche, 
P. Gegenwart, T. Eschmann, F. L. Buessen, S. Trebst, P. H. M. van Loosdrecht, 
J. van den Brink, and M. Gr\"uninger, 
\textit{Fingerprints of Kitaev physics in the magnetic excitations of honeycomb iridates}, 
Phys. Rev. Research \textbf{2}, 043094 (2020).

\bibitem{Minola15}
M. Minola, G. Dellea, H. Gretarsson, Y. Y. Peng, Y. Lu, J. Porras, T. Loew, F. Yakhou, N. B. Brookes,
Y. B. Huang, J. Pelliciari, T. Schmitt, G. Ghiringhelli, B. Keimer, L. Braicovich, and M. Le Tacon,
\textit{Collective Nature of Spin Excitations in Superconducting Cuprates Probed by Resonant
	Inelastic X-Ray Scattering},
Phys. Rev. Lett. \textbf{114}, 217003 (2015).

\bibitem{Kim12}
J. Kim, D. Casa, M.H. Upton, T. Gog, Y.-J. Kim, J.F. Mitchell, M. van Veenendaal, M. Daghofer,
J. van den Brink, G. Khaliullin, and B.J. Kim,
\textit{Magnetic Excitation Spectra of Sr$_2$IrO$_4$ Probed by Resonant Inelastic X-Ray Scattering:
	Establishing Links to Cuprate Superconductors},
Phys. Rev. Lett. \textbf{108}, 177003 (2012).

\bibitem{Kim14}
J. Kim, M. Daghofer, A.H. Said, T. Gog, J. van den Brink, G. Khaliullin, and B.J. Kim,
\textit{Excitonic quasiparticles in a spin-orbit Mott insulator},
Nature Commun. \textbf{5}, 4453 (2014).

\bibitem{footnotePearson}
The Pearson VII function allows one to smoothly change the line shape from Lorentzian to Gaussian
\cite{Wang05Pearson}, taking into account both the intrinsic line width and the instrumental
energy resolution. Compared to the Voigt function, the Pearson VII function drastically
simplifies the fitting process.

\bibitem{Wang05Pearson}
H. Wang and J. Zhou,
\textit{Numerical conversion between the Pearson VII and pseudo-Voigt functions},
J. Appl. Cryst. \textbf{38}, 830 (2005).

\bibitem{Perkins14}
N.B. Perkins, Y. Sizyuk, and P. W\"{o}lfle, 
\textit{Interplay of many-body and single-particle interactions in iridates and rhodates},
Phys. Rev. B \textbf{89}, 035143 (2014).

\bibitem{KimKhalNa14}
B.H. Kim, G. Khaliullin, and B.I. Min, 
\textit{Electronic excitations in the edge-shared relativistic Mott insulator: Na$_2$IrO$_3$}, 
Phys. Rev. B \textbf{89}, 081109(R) (2014).

\bibitem{YingLi}
Y. Li, S.M. Winter, and R. Valent{\'\i}, private communication.

\bibitem{Yadav18}
R. Yadav, R. Ray, M. S. Eldeeb, S. Nishimoto, L. Hozoi, and J. van den Brink,
\textit{Strong Effect of Hydrogen Order on Magnetic Kitaev Interactions in H$_3$LiIr$_2$O$_6$},
Phys. Rev. Lett. \textbf{121}, 197203 (2018).

\bibitem{Li18}
Y. Li, S. M. Winter, and R. Valent{\'\i},
\textit{Role of Hydrogen in the Spin-Orbital-Entangled Quantum Liquid Candidate H$_3$LiIr$_2$O$_6$},
Phys. Rev. Lett. \textbf{121}, 247202 (2018).

\bibitem{Knolle19b}
J. Knolle, R. Moessner, and N. B. Perkins,
\textit{Bond-Disordered Spin Liquid and the Honeycomb Iridate H$_3$LiIr$_2$O$_6$:
	Abundant Low-Energy Density of States from Random Majorana Hopping},
Phys. Rev. Lett. \textbf{122}, 047202 (2019).

\bibitem{MorettiSr3}
M. Moretti Sala, V. Schnells, S. Boseggia, L. Simonelli, A. Al-Zein, 
J.G. Vale, L. Paolasini, E.C. Hunter, R.S. Perry, D. Prabhakaran, A.T. Boothroyd, 
M. Krisch, G. Monaco, H.M. R{\o}nnow, D.F. McMorrow, and F. Mila, 
\textit{Evidence of quantum dimer excitations in Sr$_3$Ir$_2$O$_7$}, 
Phys. Rev. B \textbf{92}, 024405 (2015).

\bibitem{He16}
H. He, A.X. Gray, P. Granitzka, J.W. Jeong, N.P. Aetukuri, R. Kukreja, 
L. Miao, S.A. Breitweiser, J. Wu, Y.B. Huang, P. Olalde-Velasco, J. Pelliciari, W.F. Schlotter, 
E. Arenholz, T. Schmitt, M.G. Samant, S.S.P. Parkin, H.A. D\"{u}rr, and L.A. Wray, 
\textit{Measurement of collective excitations in VO$_2$ by resonant inelastic x-ray scattering},
Phys. Rev. B \textbf{94}, 161119(R) (2016). 

\bibitem{Schuelke11}
W. Sch\"{u}lke and C. Sternemann, 
\textit{Charge excitations in stripe-ordered La$_{5/3}$Sr$_{1/3}$NiO$_4$ and 
La$_{15/8}$Ba$_{1/8}$CuO$_4$: Interpretation of the anomalous momentum transfer 
dependence via fluorescence interferometry}, 
Phys. Rev. B \textbf{84}, 085143 (2011). 


\end{thebibliography}
\end{document}